# Two-colour coherent control of nuclear and electron dynamics in photoionization of molecular hydrogen with FEL pulses

**Authors:** F. Holzmeier[1,2,†,*], A. Gonzalez-Castrillo[3,4], T. M. Baumann[5], R. Y. Bello[6], C. Callegari[7], M. Di Fraia[7], M. Lucchini[2,8], M. Meyer[5], O. Plekan[7], K. C. Prince[7,9], E. Roussel[10], R. Wagner[5], F. Martín[3,4,*], A. Palacios[3,11*], D. Dowek[1*]

**Affiliations:**
[1]Université Paris-Saclay, CNRS, Institut des Sciences Moléculaires d'Orsay, 91405 Orsay, France.
[2]Dipartimento di Fisica, Politecnico di Milano, 20133 Milan, Italy
[3]Departamento de Química, Universidad Autónoma de Madrid (UAM), 28049 Madrid, Spain
[4]Instituto Madrileño de Estudios Avanzados en Nanociencia (IMDEA-Nano), Campus de Cantoblanco, 28049 Madrid, Spain
[5]European XFEL, 22869 Schenefeld, Germany
[6]Departamento de Química Física Aplicada, UAM, 28049 Madrid, Spain
[7]Elettra-Sincrotrone Trieste, 34149 Basovizza, Italy
[8]Institute for Photonics and Nanotechnologies, IFN-CNR, 20133 Milano, Italy
[9]Department of Surface and Plasma Science, Charles University, Prague 18000, Czech Republic
[10]Université Lille, CNRS, UMR 8523, PhLAM-Physique des Lasers Atomes et Molécules, 59000 Lille, France
[11]Condensed Matter Physics Center (IFIMAC), Universidad Autónoma de Madrid (UAM), 28049 Madrid, Spain

*Correspondence to: fabian.holzmeier@imec.be, fernando.martin@uam.es, alicia.palacios@uam.es, danielle.dowek@universite-paris-saclay.fr
[†] Present address: IMEC, 3001 Leuven, Belgium

**Abstract**

Extension of coherent ω-2ω control scheme recently implemented in free-electron lasers (FEL) to molecular systems offers new opportunities to control chemical dynamics on the electronic timescale, potentially allowing for the steering of reactions along previously inaccessible pathways. We have implemented such a scheme at the seeded FERMI FEL to retrieve the relative phases between one-photon (frequency 2ω) and two-photon (frequency ω) ionization paths in the hydrogen molecule as a function of photoelectron energy and emission angle. The narrow bandwidth of the XUV pulses enables selective excitation of vibrational levels of neutral intermediate $H_2$ states in the two-photon ionization path. Here we focus on ω-2ω ionization of $H_2$(X $^1\Sigma_g^+$, $v$=0) into the $H_2^+$(X $^2\Sigma_g^+$, $v_f$) ground state involving the $H_2$(B $^1\Sigma_u^+$, $v'$=6) intermediate state. The relative phases of the ω and 2ω interfering photoionization amplitudes exhibit a strong dependence on photoelectron energy, i.e., on the final vibrational state $v_f$ in the $H_2^+$ cation. With the help of accurate theoretical calculations, the observed phase jumps are assigned to the coupled electronic and nuclear dynamics at play in the two-photon process, significantly influenced by $H_2$($^1\Sigma_g^+$ and $^1\Pi_g$) autoionizing states and the mapping of the $H_2$(B $^1\Sigma_u$, $v'$=6) intermediate-state nuclear wavefunction into the final vibrational states of $H_2^+$(X $^2\Sigma_g^+$). The present work establishes the fundamental concepts required to access coupled electron–nuclear dynamics in molecules using ω–2ω coherent control schemes currently available at free-electron laser facilities.

## Main Text

## Introduction

Controlling the dynamics of a chemical reaction on the time scale of electron and nuclear motions, i.e. atto- and femtoseconds, provides the prospect of steering its outcome into yet unexplored paths. High-power ultrafast table-top lasers can nowadays produce pulses with such durations, so that chemical reactions can be studied and controlled in the time domain, typically employing pump-probe techniques.[1–4]

Accessing chemical reactions with attosecond time resolution ideally requires coherent light sources, but not necessarily attosecond pulses[5–9]. Several experimental approaches have been designed to date, using two coherent collinearly polarized light fields, with typically the fundamental (frequency ω) and a higher harmonic ($n$ω), to steer electron dynamics by controlling their ω-$n$ω relative optical phase[7,10–12]. The non-linearity of this process requires a high intensity of the fundamental field, which is challenging for pulses in the extreme-ultraviolet (XUV) or X-ray range, but whose optical phase control has been demonstrated to reach resolution of a few-attoseconds[7–9]. In particular, the seeded free electron laser (FEL) FERMI[13] has been shown to provide pulses in the XUV that are both sufficiently intense and possess a high degree of longitudinal coherence to set up coherent ω-2ω control schemes for the photoionization of atoms, in which a one-photon (frequency 2ω) ionization pathway interferes with a two-photon (frequency ω) process involving a resonant intermediate state[7,10]. Such interferences between one-photon ionization (OPI) and two-photon ionization (TPI) pathways, leading to photoelectrons with opposite parities, *ungerade* ($u$) and *gerade* ($g$), are observed in the photoelectron angular distributions (PADs) as an asymmetry with respect to the plane perpendicular to the light polarization axis, which depends on the relative phase between the ω and 2ω fields[7,10,14–16]. Inspired by the pioneering studies on neon and helium[7,10], several evolutions at FERMI showed that experiments utilizing two coherent FEL pulses to ionize atomic targets can be used to determine the absolute phase relationship between the fundamental and the second harmonic,[12] to extract phase differences between two-photon ionization channels without angular resolution[17], or angle resolved phase differences between one-photon and two-photon ionization pathways[18], to reveal the role of resonant states in photoemission,[19,] or to achieve attosecond coherent control of electronic wave packets.[20] So far, similar experimental developments have not been reported in molecular systems.

Coherent control in molecules differs from the atomic case in two main aspects: the additional degrees of freedom due to the nuclear motion characterized by the rovibrational structure of the neutral and ionic states, and the enhanced number of partial waves describing the electronic wavefunction due to the breaking of spherical symmetry of the target. The significant dependence of one-photon ionization phases on the nuclear degrees of freedom in molecules has been investigated recently e.g., in RABBITT (reconstruction of attosecond beating by two-photon transitions) XUV-pump IR-probe experiments[21–23], and predicted to lead to asymmetric photoelectron emission both in the laboratory and molecular frames[24]. In contrast to RABBITT and other control schemes based on attosecond chronoscopic techniques, where the two-photon coherent pathways involve the absorption or emission of an IR photon within the electronic continuum - thereby introducing an additional continuum-continuum phase which can be particularly challenging to quantify in molecular targets[25] - the ω-2ω scheme enables the direct extraction of the phase difference between the one-photon (2ω) and two-photon (ω) ionization amplitudes. Moreover, the very narrow bandwidth of the employed pulses and the involvement of a single total photon energy ensure high spectral resolution, significantly reducing - or even eliminating - the problem of vibrational and electronic congestion that is intrinsic to molecular systems.

Here, we present a combined experimental and theoretical study of time (optical phase) resolved quantum interferences between one-photon and two-photon paths captured in the angle-and energy-resolved single ionization of randomly oriented molecules. We choose as a prototype reaction single ionization of the hydrogen molecule in its ground state $H_2(X\ ^1\Sigma_g^+, v=0)$ leading to a vibrationally excited hydrogen molecular cation in its ground electronic state $H_2^+(X\ ^2\Sigma_g^+, v_f)$ and a photoelectron whose energy distribution extends across a 2.5-eV range[26], where the two-photon pathway involves resonant excitation of a well-defined intermediate vibronic state of the neutral $H_2$ molecule. We apply the coherent ω-2ω two-colour scheme available at FERMI and we further exploit the high spectral resolution and precise frequency tunability to be in resonance with selected $v'$ levels of the $H_2(B\ ^1\Sigma_u^+, v')$ intermediate state in the TPI path (a particular case of resonance enhanced multiphoton ionization, REMPI), which both modulates the range of internuclear distances R from which the second photon accesses the electronic continuum[27], and restricts the range of accessible molecular orientations due to the cylindrical symmetry of the molecule. Photoelectron velocity map imaging (VMI) at the Low Density Matter (LDM) end-station[28,29] provides the necessary energy resolution to capture two interfering quantum pathways leading to the same vibrational level of the final ionic state $H_2^+(X\ ^2\Sigma_g^+, v_f)$, and the capability to record the corresponding photoelectron angular distributions. For the studied photoionization of the hydrogen molecule, the measured PADs demonstrate the clear signature of photoelectron emission asymmetry relative to a plane perpendicular to the polarization axis as described above for atoms, resulting from the interference between the OPI and TPI channels involving continuum molecular states with *u* and *g* symmetry and populating the same final vibronic state. The asymmetry, characterized by the $\beta_1$ and $\beta_3$ anisotropy parameters in the Legendre polynomial expansion of the PAD, oscillates with the optical phase shift between the ω and 2ω fields, giving access to the energy- and angle-dependent relative phases between the OPI and TPI amplitudes. The experimental results are well predicted by ab initio simulations performed within the second order time-dependent perturbation theory accounting for the random orientation of the target, which further allow for disentangling the symmetry of the contributing channels (see Methods). The angle resolved relative phases between the TPI and OPI amplitudes reveal a strong dependence on the photoelectron energy $\varepsilon$, i.e., on the final vibrational state populated in the $H_2^+$(X $^2\Sigma_g^+$, $v_f$) cationic ground state. They carry information on both the electronic dynamics involving autoionizing states of $g$ symmetry, and the nuclear dynamics within the intermediate $H_2$(B $^1\Sigma_u^+$, $v'$) resonant states projected onto the $H_2^+$(X $^2\Sigma_g^+$, $v_f$) final states.

The results of this benchmark study show how the additional control knobs provided by molecular systems can be used when applying ω-2ω coherent control schemes to extract dynamical information with sub-fs resolution by using a seeded FEL.

## Results

### Molecular Interferometer employing ω-2ω FEL pulses in the VUV.

The possibility of achieving active control of a molecular photoionization process in an $\omega$-$2\omega$ scheme relies on the feasibility of establishing both the two-photon and one-photon ionization pathways and to characterize the conditions where interferences between them may occur. These are sketched in Fig. 1 for the chosen photon energy range, around $\hbar\omega$ = 12.1 eV.

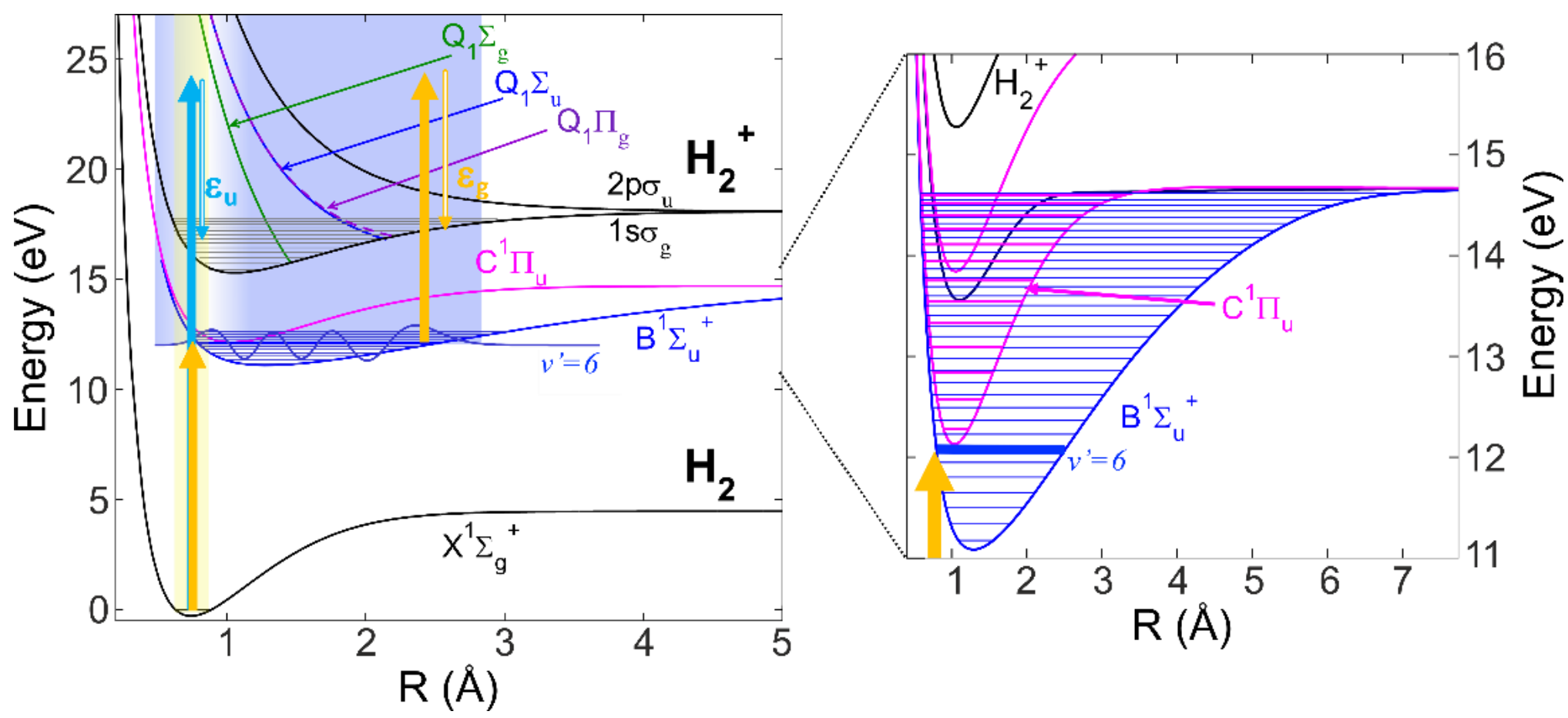

**Fig. 1 Experimental scheme and potential energy curves of the relevant electronic states of $H_2$ and $H_2^+$.** Left panel: The one-photon ionization pathway from the $H_2$ (X$^1\Sigma_g^+$, $v$=0) ground state is depicted in blue for a photon energy $\hbar 2\omega$ = 24.2 eV, while the yellow arrows indicate a two-photon pathway which resonantly populates the $v'$=6 level of the B $^1\Sigma_u^+$ intermediate

state using photons of energy $\hbar\omega$ = 12.1 eV. The narrow light-yellow shaded area (left) shows the Franck-Condon (FC) region of the neutral ground state, whereas the wider blue-shaded area highlights the internuclear distance range of the $v'$=6 vibrational wavefunction of the B state. The yellow and blue thin arrows indicate emission of electron partial waves $nl\lambda_u$ of *ungerade* ($\varepsilon_u$) and $nl\lambda_g$ of *gerade* ($\varepsilon_g$) symmetry for the one- and two-photon process, respectively. The lowest autoionizing states that can be populated at this total photon energy are shown as well[30,31]. Right panel: Detailed representation of the potential curves describing absorption of the first ω photon in the TPI process.

Starting from the $H_2(X\ ^1\Sigma_g^+, v=0)$ initial ground state, we focus on the dominant one- and two-photon absorption pathways leading to $H_2^+$ in its ground electronic state and vibrational state $v_f$, $H_2^+(X\ ^2\Sigma_g^+, v_f)$. This scheme includes: (*i*) ionization with a single $\hbar 2\omega$ photon (blue arrow in Fig. 1) absorbed at small internuclear distance, about $R_e \approx 0.7$ Å, from the $H_2(X\ ^1\Sigma_g^+, v=0)$ ground state, (*ii*) two-photon ionization via any (virtual or resonant) intermediate vibronic $^1\Sigma_u^+$ or $^1\Pi_u$ state, where one $\hbar\omega$ photon is first absorbed at internuclear distance $R_e$. For the selected photon energy, $\hbar\omega$ = 12.1 eV, REMPI takes place via the specific vibrational intermediate state $H_2(B\ ^1\Sigma_u^+, v'=6)$ (yellow arrows in Fig. 1), which is ionized after absorption of a second $\hbar\omega$ photon in the wide range of internuclear distances $R_e \leq R \leq R_M$ ($R_M \approx 3$ Å) due to the shallow potential energy curve of the B state, as reported earlier for two-photon ionization of the $H_2(B\ ^1\Sigma_u^+, v'=9)$ state[27]. As a consequence of the Σ symmetry of the intermediate state, the neutral target undergoing the interference between OPI and TPI is characterized by a distribution of molecular orientations relative to the light polarization axis with a molecular anisotropy parameter $\beta_{Molec} = 2$. Pathway (*i*) populates a continuum state of *u* symmetry, while the final continuum state for (*ii*) has *g* symmetry. It is noteworthy that for the chosen photon energies autoionizing states of *u* symmetry are not significantly populated after one-photon absorption since they lie at higher energies in the FC region[31,32]. By contrast, autoionizing states[31] of *g* symmetry, in particular the lowest $Q_1(^1\Sigma_g^+)$ state, have been predicted to play a role in two-photon ionization, influencing notably the vibrational distribution of the $H_2^+(X\ ^2\Sigma_g^+, v_f)$ final state.[27,33]

The one-photon photoelectron spectrum describing direct photoionization into the $H_2^+(X\ ^2\Sigma_g^+, v_f)$ ground electronic state in the neighbourhood of the studied $\hbar 2\omega$=24.2 eV photon energy has been well characterized in earlier high resolution studies[34]. Figure 2a shows the measured photoelectron spectra obtained by ionizing a sample of randomly oriented $H_2$ molecules in the two-photon process, while scanning the energy of the $\omega$ beam between $\hbar\omega$ = 12.025 and 12.4 eV. When the photon energy matches a specific vibrational level of the B $^1\Sigma_u^+$ state, namely $v'$=6, 7, 8 spaced by about 140 meV[35] , the REMPI leads to a maximum in the photoelectron yield[27,36] demonstrating that the bandwidth of the 50-fs pulses delivered by FERMI enables a clear selection of single vibrational levels in the intermediate state (i.e., the pulse duration is longer than the vibrational period associated with the $v'$ levels explored in Fig. 2, of the order of $\tau_{vib} \approx$ 30 fs for $v$ = 6)[37]. Since the experiment was performed with randomly oriented molecules, the simulations must account for all optically allowed molecular states accessible from the $H_2(X\ ^1\Sigma_g^+)$ ground state by one- and two-photon transitions (bound and continuum electronic and vibrational states) with linearly polarized light. In contrast with our previous calculations[27] that assumed fixed in space molecules parallel to the polarization axis, the multiphoton ionization of the randomly oriented target now includes contributions from all possible orientations. The measured spectrum is in excellent agreement with the predictions of these time-dependent second-order perturbation theory, which have been convoluted with a normalized Gaussian function of width $\sigma \approx$ 110 meV to account for the experimental photoelectron energy resolution (Fig. 2b). Figure 2a also captures the REMPI process through the C $^1\Pi_u(v'=0)$ state reached at ~12.32 eV, which has been confirmed by theory, although with a much smaller signal, and is thus hardly visible in panel 2b.

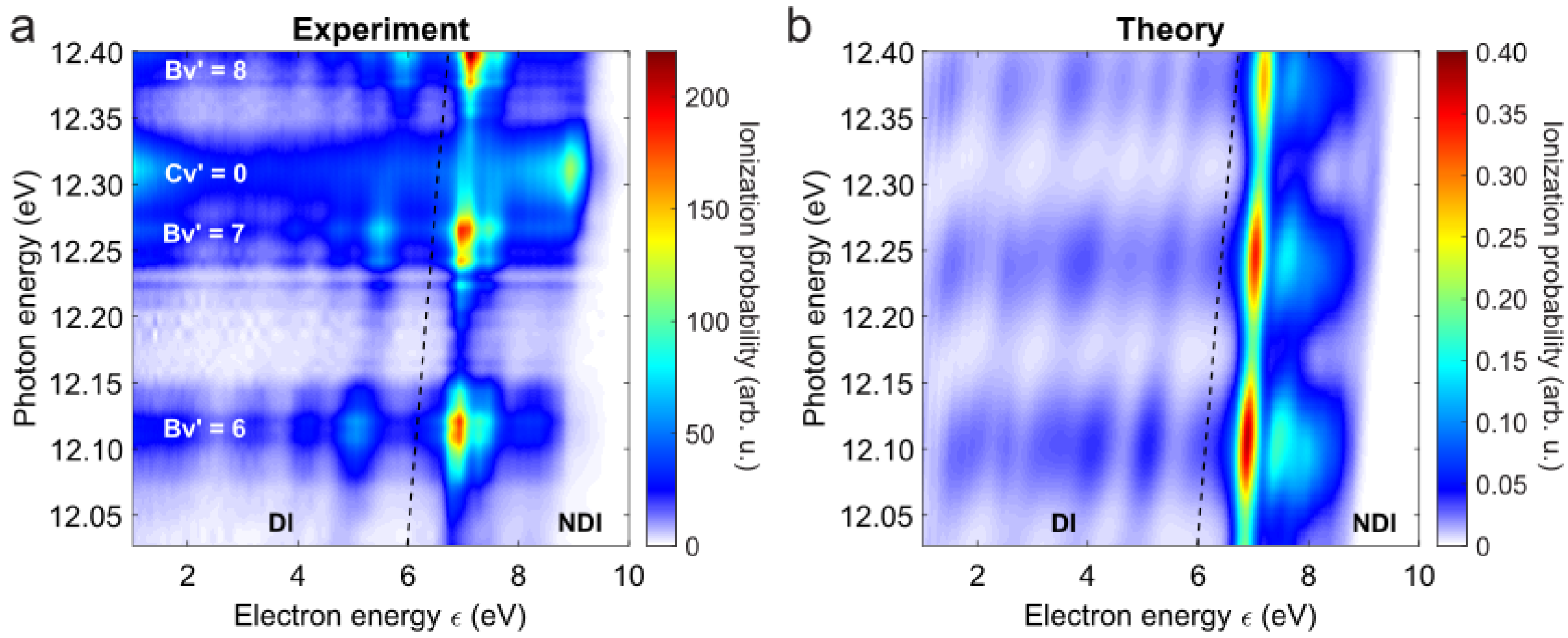

**Fig. 2 Photoelectron spectra for resonantly enhanced two-photon ionization of $H_2$.** (a) Experimental photoionization yield as a function of the electron kinetic energy $\varepsilon$ and the photon excitation energy $\hbar\omega$ for two-photon ionization of $H_2$. The spectrum shows that the 50-fs pulses of the Fermi FEL enable the selection of a single vibrational level in the resonant B $^1\Sigma_u^+$ intermediate electronic state (clear structures appear at those photon energies in resonance with the $v'$=6, $v'$=7 and $v'$=8 vibrational states of the $H_2$(B $^1\Sigma_u^+$, $v'$) intermediate state, denoted as B$v'$=6, B$v'$=7 and B$v'$=8). The structure denoted as C$v'$=0 corresponds to resonant excitation of the $v'$=0 vibrational state associated with the C $^1\Pi_u$ intermediate electronic state. The dotted line shows the limit between non-dissociative ionization (NDI) and dissociative ionization (DI) (b) Theoretical photoionization spectra, convoluted with a Gaussian function of 110 meV width to account for the experimental photoelectron energy resolution (see text).

The experimental and theoretical two-photon photoelectron spectra (PES) for the photon energy of $\hbar\omega$ = 12.1 eV involving resonant excitation of the $H_2$(B $^1\Sigma_u^+$, $v'$=6) state, displayed in more detail in Figs. 3a and 3b, respectively (red curves), are in very good agreement with each other. They are qualitatively similar to those observed previously for the B($v'$=9) state reached with $\hbar\omega$ =12.51 eV using 100-fs pulses at FERMI[27], further supporting the interpretation of the REMPI process illustrated in Fig 1. This explains the unusual and significant enhancement of dissociative ionization (DI) relative to non-dissociative ionization (NDI) in the two-photon ionization process as compared to one-photon ionization.[27,36,38] DI is associated with the photoelectron energy range $0 \leq \varepsilon \lesssim 6.1$ eV, and NDI with $6.1 \leq \varepsilon \lesssim 8.8$ eV, respectively, for $\hbar\omega$ = 12.1 eV (see Figs. 3a and 3b, red curves). For DI, the observed oscillatory structure in the PES originates mainly from the projection of the vibrational wavefunction of the B $^1\Sigma_u^+$($v'$) intermediate state onto the repulsive potential of the dissociative $H_2^+(2p\sigma_u)$ continuum[27]. For NDI the overall structure of the PES is predominantly influenced by the FC factors between the B $^1\Sigma_u^+(v'=6)$ state with the $H_2^+(X\ ^2\Sigma_g^+, 1s\sigma_g, v_f)$ bound states of the cation (see Fig. 3d). From an inspection of the vibrational wavefunctions in the intermediate and final states (see Fig. S1 in the Supplementary Material, SM), the pronounced maximum around the $v_f$=8 and $v_f$ =9 levels ($\varepsilon \approx$ 6.8-7 eV for $\hbar\omega$ =12.1 eV) is attributed to absorption of the second photon mainly at large internuclear distances corresponding to the maximum of the $v'$=6 nuclear wave-function at the outer turning point, with a clear minimum for $v_f$ =7 ($\varepsilon \approx$ 7.2 eV). Absorption of the second photon at small values of R (secondary maximum of the wavefunction at the inner turning point) reaches the continuum closer to the FC region of the $H_2(X\ ^1\Sigma_g^+, v=0)$ ground state, favouring population of the $v_f$ =3 and $v_f$ =2 levels ($\varepsilon \approx$ 8-8.2 eV). While the FC factors in Fig. 3d allow one to explain the overall profile, the R-dependencies of the dipole moment and the coherent contribution of autoionizing states further determine the PES[33,39]. We note that the observed PES favours a larger probability into the $v_f$ =9 level, confirmed by the calculated PES with and without doubly excited states reported in Fig. S2 (SM) prior to convolution with the 100-meV-width Gaussian function.

The one-photon PES (blue curves in Figs. 3a and 3b) induced by the $2\omega$ ionization is largely dominated by NDI into the $H_2^+(X\ ^2\Sigma_g^+, 1s\sigma_g, v_f)$ ground state of the cation,[34] with a contribution of only a few percent from DI[40]. At the present photon energy corresponding to direct ioniza-tion, it is very well predicted by the FC factors between the vibrational wavefunctions of the $H_2(X\ ^1\Sigma_g^+, v=0)$ neutral state and the $H_2^+(X\ ^2\Sigma_g^+, 1s\sigma_g, v_f)$ vibronic states displaying a maximum at $v_f$=2 ($\varepsilon \approx$ 8.2 eV for $\hbar 2\omega$=24.2 eV). The electron yield decreases rapidly for higher $v_f$ levels, i.e., lower kinetic energies, up to the dissociation limit ($E_{diss}$=18.1 eV, $\varepsilon$ = 6.1 eV) as shown in Fig. 3c[34], extending to DI as an exponentially decaying tail up to $\varepsilon \approx$ 5 eV[41].

The maximum overlap of the electron yields resulting from one- and two-photon ionization therefore lies in the photoelectron energy range between 6 and 9 eV, corresponding to NDI into the bound $H_2^+(X\ ^2\Sigma_g^+, 1s\sigma_g, v_f)$ ionic states. The structure of the measured PES obtained in the $\omega$-$2\omega$ two-colour scheme is well predicted by the calculations by using a larger photon intensity for the one-photon absorption process than for the two-photon one, as shown in Figs. 3a and 3b (black curves). Despite being different from the intensities estimated during the experiment, whose optimization also involves other parameters (see Experimental Setup in Methods), we note that they have no effect on the phase variation discussed below provided that first order and second order perturbation theory hold for the OPI and TPI process, respectively, as is indeed the case (see Theory in Methods). In Fig. 3b, the computed contributions of every OPI and TPI path of $\Lambda_{u,g}$ symmetries are also shown. In the TPI process ($\omega$), the first photon is not energetic enough to reach the $\Pi_u$ intermediate states, while it resonantly excites the B $^1\Sigma_u^+(v'=6)$ state, which is accessible from the ground state only through the light components parallel to the molecular axis. The second photon absorption can now reach two different final $\Lambda_g$ symmetries in the continuum. However, because of the implicit molecular alignment in the absorption of the first photon, the $(\Sigma_g \rightarrow \Sigma_u \rightarrow \Sigma_g)$ pathway involving two parallel transitions dominates, whereas the $(\Sigma_g \rightarrow \Sigma_u \rightarrow \Pi_g)$ pathway involving a perpendicular transition at the second step has a much lower probability, as demonstrated in Fig. 3b. Both channels show a pronounced maximum for an electron energy lying between 6.8 and 7 eV, mainly governed by the FC factors from the $v'=6$ vibrational state. The contribution of the $(\Sigma_g \rightarrow \Sigma_u \rightarrow \Pi_g)$ pathway into lower $v_f$ vibrational states becomes negligible compared with that of the $(\Sigma_g \rightarrow \Sigma_u \rightarrow \Sigma_g)$ pathway, which supports the role of autoionizing states of $\Sigma_g$ symmetry in that component of the TPI process. Although the $(\Sigma_g \rightarrow \Pi_u)$ pathway is dominant in the OPI ($2\omega$) process for randomly oriented molecules, the contribution of the $(\Sigma_g \rightarrow \Sigma_u)$ pathway is still important to properly describe the interference between the TPI and OPI, as TPI mainly leads to final states of Σ symmetry.

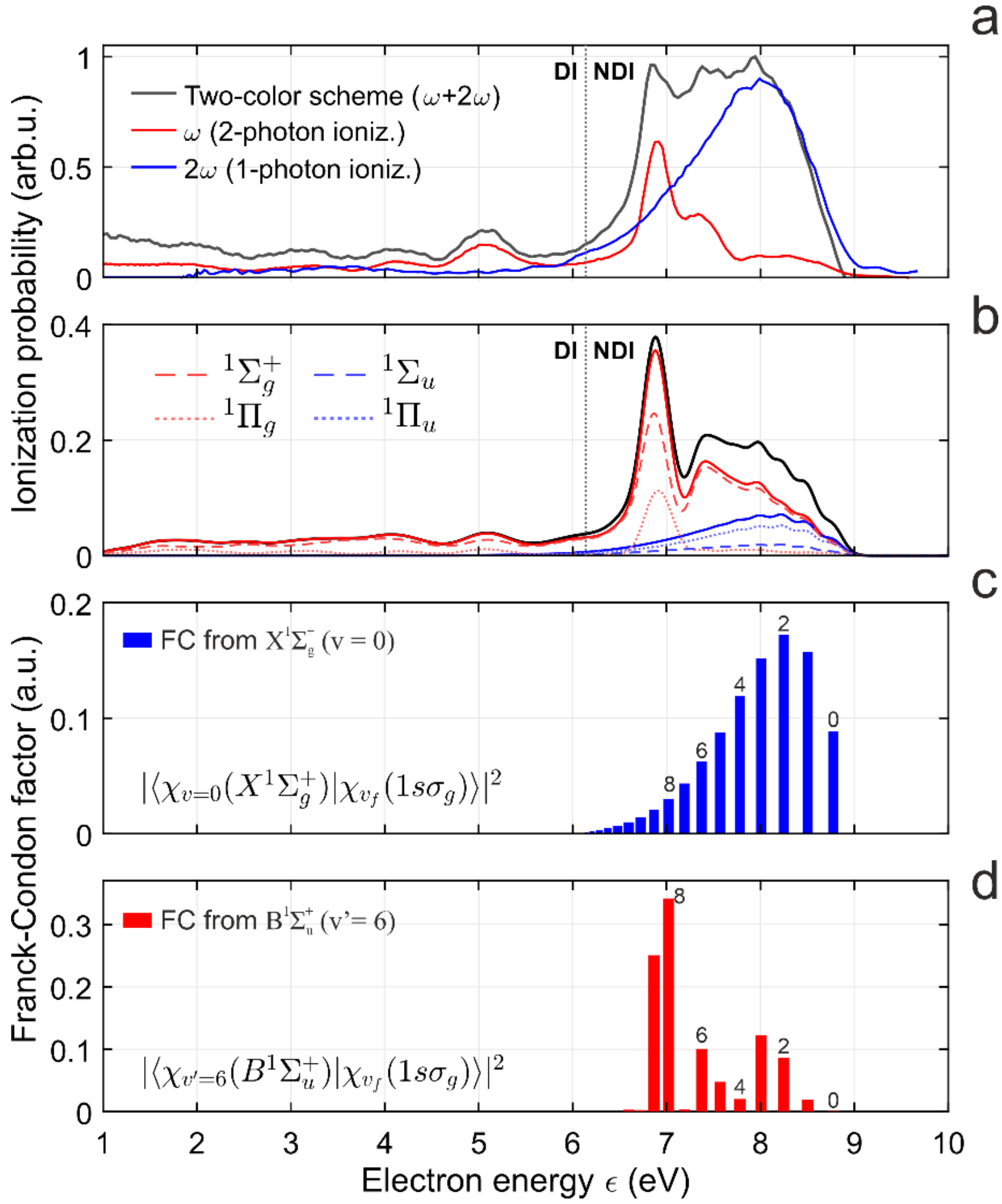

**Fig. 3. Photoelectron spectra and Franck-Condon factors for the one-photon and two-photon ionization pathways.** (a) Experimental PES profiles for one-photon (blue curve), two-photon (red curve), and the ω-2ω scheme (black curve) for ionization of $H_2$ with $\hbar\omega$ = 12.1 eV and $\hbar 2\omega$ = 24.2 eV, normalized such that the shape of the effective PES in the ω-2ω scheme is well reproduced. The limit between dissociative (DI) and non-dissociative ionization (NDI) is indicated with the vertical dashed black line. (b) Computed PESs with the same legend as in (a) for OPI (blue curve), TPI (red curve), and the ω-2ω scheme (black curve): laser peak intensities in the calculation were chosen as 2.5x$10^{13}$ W/cm$^2$ and $10^{12}$ W/cm$^2$, for the 2ω and ω FEL radiation, respectively. The theoretical spectra are also decomposed into the contributions of the different final states of Σ (thick dashed line) or Π (dotted line) symmetry for randomly oriented molecules in the ground state. (c) Computed Franck-Condon factors between the $v$=0 vibrational state in the X $^1\Sigma_g^+$ electronic state and the whole progression of vibrational states $v_f$ associated with the X $^2\Sigma_g^+(1s\sigma_g)$ state of the $H_2^+$cation. (d) Computed Franck-Condon factors between the vibrational state in resonance with the first photon absorption ($v'$=6 vibrational state in the B $^1\Sigma_u^+$ electronic state) and the progression of $v_f$

vibrational states associated with the $H_2^+$(X $^2\Sigma_g^+$(1s$\sigma_g$)) state. For illustration, selected plots of these vibrational wave functions are given in Fig. S1 (SM).

**Relative electronic phases from interference of one- and two-photon pathways in $H_2$.**

In order to implement the $\omega$-$2\omega$ interference scheme, the first five undulators of the FERMI FEL were set to produce longitudinally-coherent, horizontally-polarized pulses of $\hbar\omega$=12.1 eV, populating the B $^1\Sigma_u^+$(v'=6) intermediate state, and the last undulator was set to produce pulses of $\hbar 2\omega$=24.2 eV. The tunable optical phase shift $\phi$ between the $\omega$ and $2\omega$ fields was varied over a range of two complete periods of the $2\omega$ field (4π) with a step of $\sim 0.2\pi$ by controlling the delay of the electron bunch before the last undulator (generating $2\omega$) with a phase shifter.[7,10] This value corresponds to a temporal step of the order of 17 as. For each value of $\phi$, we measured the PAD, $I(\theta_k, \phi, \varepsilon)$, resulting from the interference of the TPI and OPI amplitudes, where $\theta_k$ is the polar emission angle relative to the radiation polarization axis. The black curves in Figs. 3a and 3b correspond to the total PES, $I(\varepsilon)$, obtained by integrating over $\theta_k$ and $\phi$.

The general form of the $I(\theta_k, \phi, \varepsilon)$ PAD resulting from the interference of the TPI and OPI amplitudes can be written as:

$$I(\theta_k, \phi, \varepsilon) = |c_\omega(\theta_k, \varepsilon)|^2 + |c_{2\omega}(\theta_k, \varepsilon)|^2 + 2|c_\omega(\theta_k, \varepsilon)||c_{2\omega}(\theta_k, \varepsilon)| \cos\big(\phi - \Delta\zeta(\theta_k, \varepsilon)\big), \quad (1)$$

where $c_\omega(\theta_k, \varepsilon)$ and $c_{2\omega}(\theta_k, \varepsilon)$ are the two- and one-photon coherent and complex amplitudes for photoionization into a final $H_2^+$(X$^2\Sigma_g^+$, $v_f$) cationic state, respectively; $\phi = \phi_{2\omega} - \phi_\omega$ is the $\omega$-$2\omega$ relative optical phase; and $\Delta\zeta(\theta_k, \varepsilon) = \zeta_\omega(\theta_k, \varepsilon) - \zeta_{2\omega}(\theta_k, \varepsilon)$ is the difference between the phases of the $c_\omega(\theta_k, \varepsilon)$ and $c_{2\omega}(\theta_k, \varepsilon)$ amplitudes.

The PAD is commonly expressed as an expansion in terms of Legendre polynomials, $P_n$, weighted by the anisotropy parameters, $\beta_n$, and multiplied by the ionization cross section $\sigma_0$:

$$I(\theta_k, \phi, \varepsilon) = \frac{\sigma_0}{4\pi}\left[1 + \sum_{n=1}^{2n_{max}} \beta_n(\phi, \varepsilon)\, P_n(\cos\theta_k)\right]. \quad (2)$$

Since the PAD involves a two-photon process and final states with different parity are reached in the two-colour ionization scheme employed here, $n_{max} = 2$ and the sum runs over even and odd components up to the fourth order. Although the computation of these parameters is significantly more complex for molecules than for atoms due to the lack of spherical symmetry, implying several partial waves and different molecular orientations with respect to the light components (see Methods), it still underlies the same formalism as for atomic targets for each given final state, characterized here by the photoelectron energy $\varepsilon$ and the vibrational level $v_f$. Among the four $\beta_n$ parameters, we focus the analysis on the odd-rank $\beta_1$ and $\beta_3$, that fully characterize the asymmetry of the PAD with respect to the polarization axis, and form the basic observables of this study, with $b_1(\varepsilon)$, $b_3(\varepsilon)$ their amplitudes and $\eta_1(\varepsilon), \eta_3(\varepsilon)$ their phases. In the perturbative regime they exhibit a $\phi$ dependence according to:

$$\beta_{1,3}(\phi, \varepsilon) = b_{1,3}(\varepsilon)\cos\left(\phi - \eta_{1,3}(\varepsilon)\right) \quad (3)$$

while $\beta_2$ and $\beta_4$ do not depend on $\phi$.[12,18,42]

## Anisotropy parameters $\boldsymbol{\beta_1}$ and $\boldsymbol{\beta_3}$

In the experiment, $\beta$-parameters were obtained by inverting the photoelectron image for each optical phase shift $\phi$ using the pBASEX algorithm[43]. The results for $\beta_1$ and $\beta_3$, shown as bidimensional histograms in terms of $\phi$ and $\varepsilon$ in Figs. 4a and 4b for a 4π variation of the optical phase shift (two periods), demonstrate that $\beta_1$ and $\beta_3$ oscillate periodically with $\phi$ for each photoelectron energy $\varepsilon$ in the investigated 5-8.7 eV energy range. They both exhibit very similar trends, in good agreement with the computed $\beta_1$ and $\beta_3$ values plotted in Figs. 4c and 4d. The corresponding $\eta_1(\varepsilon), \eta_3(\varepsilon)$ phases (modulo 2π) and $b_1(\varepsilon)$, $b_3(\varepsilon)$ amplitudes show a non-trivial dependence on the electron energy $\varepsilon$ characterizing the vibrational levels in the $H_2^+(X^2\Sigma_g^+, v_f)$ ionic state. The $\eta_{1,3}$ phases on which we focus the discussion are represented as superimposed black curves on the $\beta_1$ and $\beta_3$ histograms, in Figs. 4a and 4b for the experiment, and Figs. 4c and 4d for theory. A $2\pi$ shift was applied to $\eta_1$ and $\eta_3$ for the experimental data at $\varepsilon \approx 6.8$ eV in order to visualize the phase variation within the chosen 4π scale.

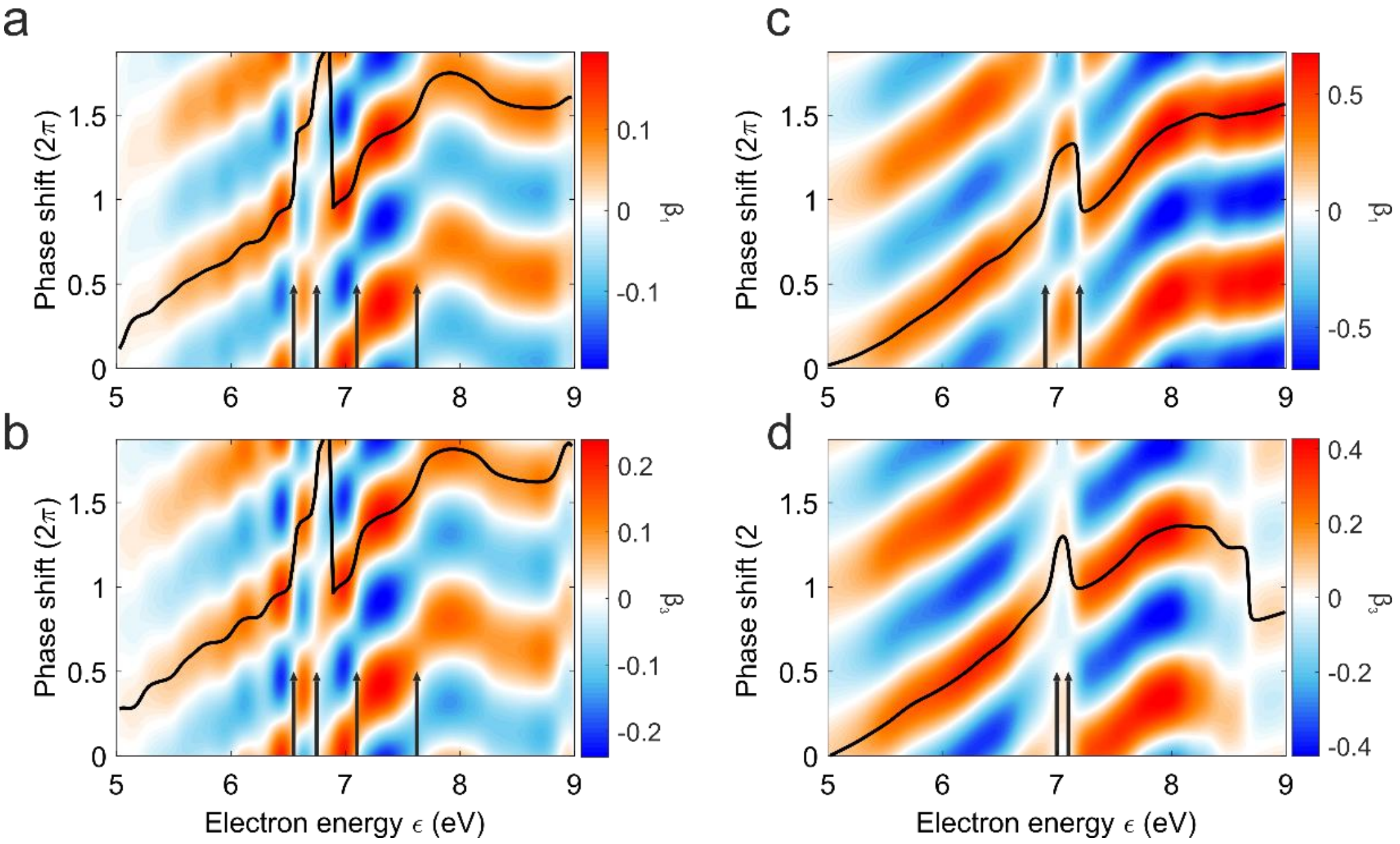

**Fig. 4: $\beta_1$ and $\beta_3$ asymmetry parameter ($\varepsilon$, $\phi$) bidimensional histograms:** (a), (b), Experiment (residual constant background component subtracted in the 2D plots); (c) (d) Calculations. $\beta_1$ (first row) and $\beta_3$ (second row) exhibit clear oscillations as a function of the optical phase shift $\phi$ for each electron energy $\varepsilon$. The 20 raw data points for each electron energy were fitted by a sinusoidal function and evaluated at 100 points, i.e., phase shift steps of $0.02 \cdot 2\pi$. $\eta_1(\varepsilon)$ and $\eta_3(\varepsilon)$ phases as defined in Eq. (3) are shown as black full line, with vertical arrows pointing to the position of the jumps (see text): a 2π shift was applied to $\eta_1(\varepsilon)$ and $\eta_3(\varepsilon)$ for the experimental data at $\varepsilon \approx 6.8$ eV in order to visualize the phase variation within the chosen 4π scale. The statistical uncertainties in extracting the phases (two standard deviations in the sinusoidal fit of the experimental data) are of the order of 1%.

The measured and computed $\eta_1$ and $\eta_3$ phases increase monotonically between 5 eV and 8 eV, with an overall slope of about 3.5 rad/eV. On top of this monotonic variation, they display pronounced phase jumps in the 6.5-8 eV energy range corresponding to $3 \leq v_f \leq 12$. The values of $\eta_1$ and $\eta_3$ derived from the experimental data are remarkably similar, featuring four jumps of the order of π for $\varepsilon \approx 6.55$ eV ($v_f \approx 11$) and $\varepsilon \approx 6.75$ eV ($v_f \approx 10$), then closer to π/2 around 7 eV≤ $\varepsilon$≤ 7.2 eV ($v_f \approx 8, 7$) and 7.5 eV≤ $\varepsilon$≤ 7.75 eV ($v_f \approx 5, 4$). In the computed phase variations, two jumps of the order of π then -π are found around $\varepsilon \approx 6.9$ and 7.2 eV ($v_f \approx 9, 7$) for $\eta_1$, and around $\varepsilon \approx 7$ eV and 7.1 eV ($v_f \approx 8, 7$) for $\eta_3$.

The remarkable similarity between the $\beta_1$ and $\beta_3$ parameters, and the $\eta_1(\varepsilon)$ and $\eta_3(\varepsilon)$ phases, in both the experiment and the theory, can only be achieved if a single term dominates in the complex expressions of $\beta_1$ and $\beta_3$ given in Eqs. (8)-(10) of Methods, the only difference between them being a simple proportionality factor. The identified dominant term is proportional to the product $c_{\Pi_u}^{l=1} c_{\Sigma_u \Sigma_g}^{l=2}$, where $c_{\Pi_u}^{l=1}$ is the amplitude for the pure perpendicular OPI path populating a $\Pi_u$ final state and $c_{\Sigma_u \Sigma_g}^{l=2}$ the amplitude for the pure parallel TPI path to a $\Sigma_g$ final state, meaning that a single partial wave plays a dominant role in the electronic continuum for TPI ($l$=2 i.e., $d\sigma_g$) as well as for OPI ($l$=1 i.e., $p\pi_u$), respectively (though not all other contributions are necessarily zero). This is the consequence of both the implicit alignment imposed by the resonantly enhanced TPI path and the dominance of the $\Pi_u$ chanel in OPI. In the calculations, the only exception to this simple behaviour is found in the interval 7 eV ≤ $\varepsilon$≤ 7.25 eV, where contributions from other partial waves become non-negligible in the TPI path.

**Relative phase between the OPI and TPI amplitudes**

Identifying the $\phi$-dependent terms in Eqs. (1) and (2) gives access to the relative phase between the TPI and OPI amplitudes, $\Delta\zeta(\theta_k, \varepsilon)$, in terms of the energy-dependent phase shifts $\eta_1(\varepsilon), \eta_3(\varepsilon)$ and magnitudes $b_1(\varepsilon)$, $b_3(\varepsilon)$ of the $\beta_1$ and $\beta_3$ parameters, such that:

$$\Delta\zeta(\theta_k, \varepsilon) = Arg\left[ b_1(\varepsilon) P_1\left(\cos\theta_k\right) e^{i\eta_1(\varepsilon)} + b_3(\varepsilon) P_3\left(\cos\theta_k\right) e^{i\eta_3(\varepsilon)} \right] \qquad (4)$$

while the amplitude $B(\theta_k,\varepsilon)$ of the oscillatory term of the PAD $I(\theta_k, \phi, \varepsilon)$, proportional to the product $|c_\omega(\theta_k,\varepsilon)||c_{2\omega}(\theta_k,\varepsilon)|$, can be written as:

$$B(\theta_k,\varepsilon)=\left|b_1(\varepsilon)P_1\left(\cos\theta_k\right)e^{i\eta_1(\varepsilon)}+b_3(\varepsilon)P_3\left(\cos\theta_k\right)e^{i\eta_3(\varepsilon)}\right| \quad (5)$$

$B(\theta_k,\varepsilon)$ shows an overall energy dependence comparable to that of $b_1(\varepsilon)$ and $b_3(\varepsilon)$, with a maximum intensity peaked at 0° in the forward (FW) emission cone (0°≤$\theta_k$ ≤ 45°), and a significant decrease for larger $\theta_k$ angles (see Fig. S3 for a 2D polar plot example) governed by the $P_1\left(\cos\theta_k\right)$ and $P_3\left(\cos\theta_k\right)$ Legendre polynomials. The energy dependence of $\Delta\zeta(\theta_k,\varepsilon)$ displayed in Fig. 5(a) is identical for all emission angles in the dominant (0°≤$\theta_k$ ≤ 45°) angular range. It is very similar to that of $\eta_1(\varepsilon)$ or $\eta_3(\varepsilon)$ with localized phase jumps of the order of π or π/2 near specific vibrational levels $v_f$, superimposed on the quasi-monotonic variation described above.

Indeed, for energies $\varepsilon$ where $\eta_1 \approx \eta_3$ labelled as $\eta$ in Eq. (6), i.e., in the whole energy range between 5 and 8 eV for the experiment, and for theory except in the narrow region 7 eV ≤ $\varepsilon$ ≤ 7.25 eV, Eqs. (4) and (5) can be written approximately as:

$$\Delta\zeta(\theta_k,\varepsilon) \simeq Arg\left[\left(b_1(\varepsilon)P_1(\cos\theta_k)+b_3(\varepsilon)P_3(\cos\theta_k)\right)e^{i\eta(\varepsilon)}\right] \quad (6)$$

and

$$B(\theta_k,\varepsilon) \simeq |b_1(\varepsilon)P_1(\cos\theta_k)+b_3(\varepsilon)P_3(\cos\theta_k)| \quad . \quad (7)$$

The angular dependence of $\Delta\zeta(\theta_k,\varepsilon)$ at a given energy $\varepsilon$ therefore features a π shift determined by the change of sign of $b_1(\varepsilon)P_1\left(\cos\theta_k\right)+b_3(\varepsilon)P_3\left(\cos\theta_k\right)$, which occurs in the 50°-70° angular range. This is illustrated, e.g., in Fig. 5(b), for the 6.4 eV ≤ $\varepsilon$ ≤ 6.7 eV (10 ≤ $v_f$ ≤12) energy band, where a sharp phase jump close to π is found in the experiment. There, no significant $\theta_k$ dependence is observed in the 0°-45° range for $\Delta\zeta(\theta_k,\varepsilon)$ at selected energies, as well as for the other measured phase jumps. On the other hand, Fig. 5(c) highlights a more complex angular behaviour of the computed $\Delta\zeta(\theta_k, \varepsilon)$ in the 0°-45° forward emission zone for energies in the range 7.1 eV ≤ $\varepsilon$ ≤ 7.22 eV ($v_f$ =8,7), where $\eta_1$ differs from $\eta_3$, which can be ascribed to the contribution of more than one partial wave in the TPI amplitude.

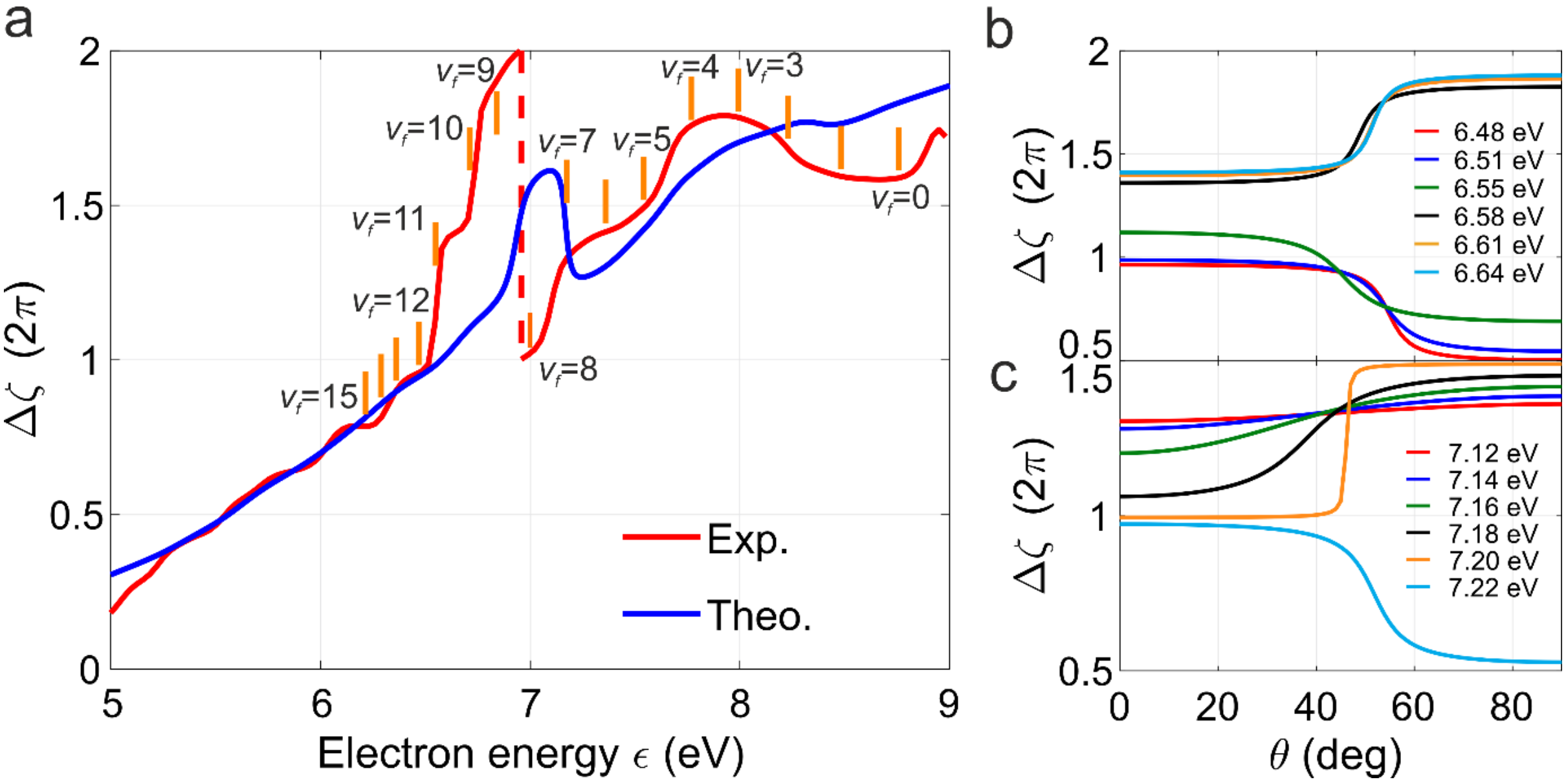

**Fig. 5 Energy and angle dependence of the measured and computed TPI-OPI relative phases $\Delta\zeta(\theta_k, \varepsilon)$.** (a) Experimental (solid) and calculated (dashed) $\Delta\zeta(\theta_k, \varepsilon)$ obtained for any emission angle in the dominant forward emission cone (0°≤$\theta_k$ ≤ 45°), as a function of $\varepsilon$, i.e. the $v_f$ level in $H_2^+$($X^2\Sigma_g^+$, $v_f$) manifold[26] (orange ticks); both series are vertically adjusted to overlap in the 5-6 eV range in order to facilitate comparison. (b) Example of measured $\Delta\zeta(\theta_k, \varepsilon)$ angular profiles at selected values of $\varepsilon$ in the interval 6.45eV ≤ $\varepsilon$ ≤ 6.74eV describing the close to π jump across the $v_f$=11 level, and (c) calculated $\Delta\zeta(\theta_k, \varepsilon)$ angular profiles for 7.10 eV ≤ $\varepsilon$ ≤ 7.30 eV describing the -0,7 π jump across the $v_f$=7 level.

Taking as a reference the $\zeta_{2\omega}$ OPI phase, which results from direct ionization at an almost fixed internuclear distance $R_e$ for all final vibrational states $v_f$ associated with the $H_2^+$(X $^2\Sigma_g^+$, $v_f$) electronic state, $\Delta\zeta(\theta_k, \varepsilon)$ provides information on the nuclear and electron dynamics at play in the TPI process. As already mentioned, dynamics in the TPI process is governed by the nuclear wave function of the intermediate $H_2$(B $^1\Sigma_u$, $v'$=6) vibronic state, and its subsequent photoionization involving the $^1\Sigma_g^+$ (and to a lesser extent $^1\Pi_g$) doubly-excited states. The main fingerprints characterizing both the measured and computed $\Delta\zeta(\theta_k, \varepsilon)$ energy profiles displayed in Fig. 5(a) for the favoured emission angles ($0° \leq \theta_k \leq 45°$), can be interpreted as follows.

The monotonic increase of $\Delta\zeta(\theta_k, \varepsilon)$ in the 5-8 eV range with the slope of 3.5 rad/eV mostly reflects the increase of the phase difference between the $l$=2 (d$\sigma_g$) and $l$=1 (p$\pi_u$) continuum wavefunctions in the TPI and OPI paths. The remarkable 6.8-7.2 eV energy band featuring ionization into $v_f = 9$ and $v_f = 8$ states corresponds to the sharp maximum of the amplitude of the TPI path involving both the $\Sigma_g$ and $\Pi_g$ ionization continua (see Fig. 3), the largest contribution of the $^1\Sigma_g^+$ continuum state ($\varepsilon$ = 6.86 eV) lying at a slightly lower electron energy than that for the $^1\Pi_g$ continuum state ($\varepsilon$ = 6.91 eV). The significant variation of the relative weight between those two final states across the pronounced band in the PES of the TPI channel results in a suddenly varying composition of the dominant partial waves of $\sigma_g$ and $\pi_g$ symmetry, which can lead to the observed $\pi$ phase jumps at about $\varepsilon \approx 7$ eV (see also Fig. 5(c)).

On the other hand, the four significant phase jumps observed in the measured $\Delta\zeta(\theta_k, \varepsilon)$, apart from the 7 eV region discussed above, for energies $\varepsilon$ corresponding, e.g., to $v_f \approx 11$, $v_f \approx 7$ or $v_f \approx 4$ (Fig.5(a)), reveal the mapping of the intermediate $v'$=6 vibrational state into the final vibrational states of $H_2^+$ populated coherently through the OPI and TPI paths, which takes place in different intervals of internuclear distances as one progresses along the $H_2^+$ $v_f$ vibrational ladder. These intervals depend on the effective overlaps between the intermediate state nuclear wavefunction and those of the $H_2^+$(X $^2\Sigma_g^+$, $v_f$) final states, and between the former and the wave functions of the $\Sigma_g$ and $\Pi_g$ autoionizing states, which depend on their lifetimes[31].

## Summary and outlook

The reported benchmark experiment and the good agreement with ab initio theory demonstrate that longitudinally coherent and tunable XUV FEL pulses available at FERMI can be applied to two-colour coherent control in photoionization of a molecule, providing the means to explore coupled electron and nuclear motions, with unique access to the nuclear wave function of a selected intermediate excited neutral state, and with atto-femtosecond time resolution.

This scheme has been applied to investigate the interference between one-photon (OPI, 2ω) and two-photon (TPI, ω) pathways for single ionization of randomly oriented $H_2(X\ ^1\Sigma_g^+, v=0)$ molecules into the $H_2^+(X\ ^2\Sigma_g^+, v_f)$ vibrational manifold, and results are reported for the photon energy $\hbar\omega$ = 12.1 eV, where the two-photon REMPI path involves resonant excitation of the $H_2(B\ ^1\Sigma_u^+, v'=6)$ intermediate vibronic state. The measured and computed $\beta_1(\phi,\varepsilon)$ and $\beta_3(\phi,\varepsilon)$ anisotropy parameters of the PAD, and in particular their phases, $\eta_1(\varepsilon)$ and $\eta_3(\varepsilon)$, are found to be remarkably similar, indicating the dominant role of a single partial wave in each pathway, a property well confirmed by the calculations except in a quite narrow energy region about $\varepsilon$ ~7 eV where more partial waves are involved. $\beta_1(\phi,\varepsilon)$ and $\beta_3(\phi,\varepsilon)$ further enabled us to access the energy- and angle-resolved relative phases $\Delta\zeta(\theta_k,\varepsilon)$ between the TPI and the OPI amplitudes for each given final state of the molecule $(\varepsilon, v_k)$. The reported coherent ω-2ω control scheme involving selection of a vibronic intermediate state can thereby be used to retrieve vibrational and electronic dynamics including their coupling and to influence the outcome of a chemical reaction with an unprecedented level of control.

A deeper insight into the coherence effects described in this work would benefit from the determination of TPI-OPI energy- and angle-dependent relative phases in the molecular frame, for each orientation of the molecular axis relative to the polarization[44]. For that purpose, future experimental work should aim at aligning the molecule with an additional external field (which is possible at FERMI[45,46]) or at making use of multi-coincidence detection methods in the dissociative ionization channel to obtain fully orientation-resolved information on the dynamics taking advantage of high repetition rates now available at XFEL facilities[47–51]. On the other hand, the use of shorter pulses, down to a few femtoseconds or even attoseconds, such as those already available at several XFEL facilities[44,48–50], would allow one to actually perform pump-probe experiments in a complementary ω/2ω scenarios[52], thus giving direct access to the coupled electron and nuclear dynamics in the absence of the external electromagnetic field, namely in between the pump and the probe pulses[33,39,52].

## Methods

### Derivation of the PAD asymmetry for the molecule:

The asymmetry of the PAD along the polarization axis is reflected in the parameters $\beta_1 = \frac{\alpha_1}{\alpha_0}$ and $\beta_3 = \frac{\alpha_3}{\alpha_0}$. When considering only the $1s\sigma_g$ final state and the non-negligible one- and two-photon paths discussed above, the expressions for $\alpha_0$, $\alpha_1$, and $\alpha_3$ are written as:

$$\alpha_0(\varepsilon) = \frac{1}{12\pi} \sum_{l=1,3,5,7} \left[ \left|c_0^l\right|^2 + \sum_{\mu=-1,1} \left|c_\mu^l\right|^2 \right] + \frac{1}{4\pi} \sum_{J=0}^{2} \frac{\langle 1,0;1,0||J,0\rangle^2}{2J+1} \left\{ \sum_{l=0,2,4,6} \left[ \langle 1,0;1,0||J,0\rangle^2 \left|c_{00}^l\right|^2 + \sum_{\mu=-1,1} \langle 1,0;1,\mu||J',\mu\rangle^2 \left|c_{0\mu}^l\right|^2 \right] \right\} \tag{8}$$

$$\alpha_1(\varepsilon) = \sum_{l=1,3,5,7} \sum_{l'=l-1,l+1} A^1(l,l') \left\{ \begin{array}{c} a_{00}^{ll'} c_0^l c_{00}^{l'} \cos\left(\theta_0^l - \theta_{00}^{l'}\right) + \\ \sum_{\mu=-1,1} \left[ \begin{array}{c} a_{0-\mu}^{ll'} c_\mu^l c_{00}^{l'} \cos\left(\theta_\mu^l - \theta_{00}^{l'}\right) + a_{\mu 0}^{ll'} c_0^l c_{0\mu}^{l'} \cos\left(\theta_0^l - \theta_{0\mu}^{l'}\right) \\ + a_{\mu-\mu}^{ll'} c_\mu^l c_{0\mu}^{l'} \cos\left(\theta_\mu^l - \theta_{0\mu}^{l'}\right) \end{array} \right] \end{array} \right\} \tag{9}$$

$$\alpha_3(\varepsilon) = \sum_{l=1,3,5,7} \sum_{l'=0,2,4,6} B^3(l,l') \left\{ \begin{array}{c} b_{00}^{ll'} c_0^l c_{00}^{l'} \cos\left(\theta_0^l - \theta_{00}^{l'}\right) + \\ \sum_{\mu=-1,1} \left[ \begin{array}{c} b_{0-\mu}^{ll'} c_\mu^l c_{00}^{l'} \cos\left(\theta_\mu^l - \theta_{00}^{l'}\right) + b_{\mu\mu}^{ll'} c_0^l c_{0\mu}^{l'} \cos\left(\theta_0^l - \theta_{0\mu}^{l'}\right) \\ + b_{\mu-\mu}^{ll'} c_\mu^l c_{0\mu}^{l'} \cos\left(\theta_\mu^l - \theta_{0\mu}^{l'}\right) + b_{\mu\mu}^{ll'} c_\mu^l c_{0-\mu}^{l'} \cos\left(\theta_\mu^l - \theta_{0-\mu}^{l'}\right) \end{array} \right] \end{array} \right\} \tag{10}$$

The amplitudes $c$ and phases $\theta$ both depend on the photoelectron energy $\varepsilon$ and are defined as follows: $c_0^{l'}, \theta_0^{l'}$ describe the parallel one-photon transition to the $\Sigma_u$ final states with components involving transition matrix elements with $\mu$=0, whereas $c_\mu^{l'}, \theta_\mu^{l'}$ (where $\mu$=-1 or 1) are used for the perpendicular one-photon path populating a $\Pi_u$ final state. The amplitudes and phases of the pure parallel two-photon pathway to $\Sigma_g$ are represented by $c_{00}^l, \theta_{00}^l$ and the two-photon pathways that are reached through a transition with parallel and perpendicular components into $\Pi_g$ final states are written as $c_{0\mu}^l, c_{0-\mu}^l$ and $\theta_{0\mu}^l, \theta_{0-\mu}^l$. The prefactors contain information about the angular components in the form of the Clebsch-Gordan coefficients and are listed explicitly in Table 1 below.

**Table 1.** Prefactors for $\alpha_1$ and $\alpha_3$ of equations (3) and (4).

| Prefactors for $\boldsymbol{\alpha_1}$ | Prefactors for $\boldsymbol{\alpha_3}$ |
|---|---|
| $\boldsymbol{A^1(l,l') = \frac{1}{2\pi}[(2l+1)(2l'+1)]^{1/2}\langle l',0;l,0\Vert 1,0\rangle}$ | $B^3(l,l') = \frac{1}{14\pi}[(2l+1)(2l'+1)]^{1/2}\langle l',0;l,0\Vert 3,0\rangle$ |
| $\boldsymbol{a_{00}^{ll'} = \frac{1}{10}\langle l',0;l,0\Vert 1,0\rangle}$ | $b_{00}^{ll'} = \frac{1}{5}\langle l',0;l,0\Vert 3,0\rangle$ |
| $\boldsymbol{a_{0-\mu}^{ll'} = -\frac{1}{45}\langle l',0;l,-\mu\Vert 1,-\mu\rangle}$ | $b_{0-\mu}^{ll'} = \sqrt{\frac{8}{75}}\langle l',0;l,-\mu\Vert 3,-\mu\rangle$ |
| $\boldsymbol{a_{\mu 0}^{ll'} = \frac{1}{45}\langle l',\mu;l,0\Vert 1,\mu\rangle}$ | $b_{\mu 0}^{ll'} = \sqrt{\frac{8}{75}}\langle l',\mu;l,0\Vert 3,\mu\rangle$ |
| $\boldsymbol{a_{\mu-\mu}^{ll'} = -\frac{1}{45}\langle l',\mu;l,-\mu\Vert 1,0\rangle}$ | $b_{\mu-\mu}^{ll'} = \frac{1}{5}\langle l,\mu;l',-\mu\Vert 3,0\rangle$ |

As visible from Eqs. (9) and (10), the argument of each cosine function for $\alpha_1$ and $\alpha_3$ is the phase difference of a combination of final states for the one- and the two-photon process, respectively.

## Experimental setup

The experiment was performed at the seeded FEL FERMI[13] at the low-density matter (LDM) beamline[28,29]. The bichromatic VUV pulse was obtained by tuning the first five undulators to the fundamental frequency $\omega$ at 12.1 eV and its sixth undulator to the second harmonic $2\omega$ at 24.2 eV. A phase shifter located between the fifth and sixth undulator delayed the electron bunch which generated the second harmonic in the sixth undulator, thereby controlling the phase with respect to the field of the first harmonic.[7,10] The intensity ratio between the two colours was optimized using a gas filter so that the two colours gave a comparable total electron yield. Typical intensities were estimated to be of the order of $0.24\times10^{12}$ W/cm$^2$ for the fundamental and in the range of 0.04-$0.22\times10^{12}$ W/cm$^2$ for the second harmonic, with transversal spot sizes measured as 0.052 mm$^2$ and 0.032 mm$^2$, respectively. The difference in spot size is partially compensated by the fact that the fundamental excites a two-photon process, so that the effective size is smaller. The duration of the XUV pulses was estimated to be about 50 fs and the pulses were focused using a Kirkpatrick-Baez arrangement to the focal position of the custom VMI-TOF spectrometer of the LDM beamline[27]. The photon beam was crossed there with a cold molecular jet of hydrogen. The phase shift between the two fields was scanned over two complete cycles in steps of $0.2\cdot\pi$ corresponding to delay steps of $\Delta\tau$ = 17.1 as. For each delay, 15000 shots (100 files with 150 shots/files) were recorded and the photoelectron images were inverted using the pBasex algorithm[43] after subtraction of the background.

**Theory**

One- and two-photon ionization amplitudes were calculated in the framework of first and second order time-dependent perturbation theory[53,54], respectively:

$$c_E^{(1)}(t) = -i \int_{t_0}^{t} dt' e^{i\left(E_{k,\epsilon,v_k} - E_{g,v_g}\right)t'} V_{E_{k,\epsilon,v_k},E_{g,v_g}}(t') \quad (11)$$

$$c_E^{(2)}(t) = -\sum_{n,v_n} \int_{t_0}^{t} dt' \int_{t_0}^{t'} dt''\, e^{i\left(E_{k,\epsilon,v_k} - E_{n,v_n}\right)t' + i\left(E_{n,v_n} - E_{g,v_g}\right)t''} V_{E_{k,\epsilon,v_k},E_{n,v_n}}(t') V_{E_{n,v_n},E_{g,v_g}}(t'') \quad (12)$$

where $E_{g,v_g}$ is the energy of the molecule in the ground electronic $g$ state and the $v_g$ vibrational state, $E_{n,v_n}$ the energy in the $n$-th electronic state and $v_n$ vibrational state, and $E_{k,\epsilon,v_k}$ the energy in the final state, which can be written as $E_{k,\epsilon,v_k} = E_{k,v_k}^{H_2^+} + \varepsilon$, with $E_{k,v_k}^{H_2^+}$ the energy of the remaining $H_2^+$ cation in the $k$-th electronic state and the $v_k$ vibrational state, and $\varepsilon$ the photoelectron energy. The sum over $n$, $v_n$ also includes the integral over the continuum part of the energy spectrum. The coupling terms are given by integrals over both electronic $\mathbf{r}$ and nuclear $R$ coordinates

$$V_{E_{k,\epsilon,v_k},E_{g,v_g}} = \int dR \int d\mathbf{r}\, \Psi_{k,\epsilon,v_k}(\mathbf{r},R) V(\mathbf{r}) \Psi_{g,v_g}(\mathbf{r},R) \quad (13)$$

$$V_{E_{k,\epsilon,v_k},E_{n,v_n}} = \int dR \int d\mathbf{r}\, \Psi_{k,\epsilon,v_k}(\mathbf{r},R) V(\mathbf{r}) \Psi_{n,v_n}(\mathbf{r},R) \quad (14)$$

$$V_{E_{n,v_n},E_{g,v_g}} = \int dR \int d\mathbf{r}\, \Psi_{n,v_n}(\mathbf{r},R) V(\mathbf{r}) \Psi_{g,v_g}(\mathbf{r},R) \quad (15)$$

where the $\Psi$ functions are eigenfunctions of the molecular Schrödinger equation, and $V$ is the laser-molecule interaction potential in the dipole approximation (velocity gauge), $V(\mathbf{r}) = \mathcal{E}(t)\boldsymbol{\epsilon} \cdot \boldsymbol{O}(\mathbf{r})$, with $\mathcal{E}(t)$ the light electric field, $\boldsymbol{\epsilon}$ the polarization direction and $\boldsymbol{O}(\mathbf{r})$ the dipole moment.

For bound electronic states, the molecular wave functions are represented in the adiabatic approximation, so that they are written as products of an electronic wave function $\psi_i^{(el)}(\mathbf{r},R)$, which is the $i$-th eigenfunction of the electronic Schrödinger equation and depends on the electronic coordinates and parametrically on the nuclear coordinates, and a nuclear wave function $\chi_{v_i}(R)$, which is the $v_i$ eigenfunction of the nuclear vibrational Schrödinger equation in the electronic state $i$:

$$\Psi_{i,v_i}(\mathbf{r},R) = \psi_i^{(el)}(\mathbf{r},R)\chi_{v_i}(R) \quad (16)$$

A similar factorization, which is even more accurate than for bound states, has been used to describe the molecular continuum states far from autoionizing resonances. When autoionizing states are populated, the adiabatic approximation does not hold any more because autoionization may take longer than the time needed by the nuclei to move significantly from their initial position. In this case, the molecular eigenfunctions adopt the more complicated form that essentially accounts for the coupling between electronic and nuclear motions by solving the full Schrödinger equation in a combined basis of both electronic and nuclear functions[31,55], within the local approximation. In our current calculations, we employ the second order time-dependent perturbation theory[27], where we have assumed that both the $\omega$ and $2\omega$ pulses have cosine square $\mathcal{E}(t)$ envelopes and a total duration of 100 fs. We first performed a fine scan of the central frequency of the $\omega$ pulse from 12 to 12.50 eV (shown in Fig. 1) for laser peak intensities of $10^{12}$ W/cm$^2$. We then carried out simulations for the two-colour $\omega$-$2\omega$ scheme using the 12.1 eV pulse with the same intensity together with a second fully overlapping pulse with 24.2 eV central frequency and a peak intensity of $2.5\times10^{13}$ W/cm$^2$. For a direct comparison with the experiment, we performed a scan of the relative phase between the two pulses from 0 to $2\pi$. From the calculated one- and two-photon amplitudes, we extracted the total, as well as the energy and angle dependent single ionization probabilities, from which we obtained the $\beta$ anisotropy parameters as discussed in the text. Simulations were performed for randomly oriented molecules, for which all energetically accessible singlet states of Σ, Π and Δ symmetries, both *gerade* and *ungerade*, were computed, and all possible one- and two-photon paths arising from all possible combinations of the three components of the dipole operator were taken into account. Dipole couplings were evaluated as described in ref.[31,56]. In addition to bound-bound couplings, all dipole couplings involving the continuum states associated with the first and the second ionization thresholds of the molecule (i.e., the electronic continua associated with the $1s\sigma_g$ and $2p\sigma_u$ states of $H_2^+$) were included. For a direct comparison with the experimental data, i.e. to reproduce the finite experimental energy resolution, the single ionization probabilities, as well as the $\beta$ parameters obtained as a function of photoelectron energies, were convoluted with a normalized Gaussian function with full-width at half-maximum (FWHM) σ=0.11 eV, defined as $(1/(\sigma\,(2\pi)^{1/2})*\exp(-(E-E_0)^2/(2\,\sigma^2))$.

Franck-Condon overlaps ($F(v,v') = \int dR\chi_v(R)\chi_{v'}^{+}(R)$) providing the FC factors in panels c and d of Fig. 3 were obtained by using the same nuclear wave functions as in the time-dependent calculations.

## Acknowledgments

The research leading to this result has been supported by the project Laserlab-Europe under Grant Agreement 654148 from the EU Seventh Framework Programme (FP7/2007-2013). The project has received funding from the European Union's Horizon 2020 research and innovation programme under the Marie Skłodowska-Curie grant agreement No 792676. This project was funded by the Ministerio de Ciencia e Innovación MICINN (Spain) through the projects PID2022-138288NB-C31 and PID2022-138288NB-C32, the 'Severo Ochoa" Programme for Centres of Excellence in R&D (CEX2020-001039-S) and the "María de Maeztu" Programme for Units of Excellence in R&D (CEX2023-001316-M), and by Comunidad de Madrid through the projects MATRIX-CM (TEC-2024/TEC- 85) and ACXIOM-CM (SYG-2024/TEC-815). All calculations were performed at the Marenostrum computer of the Barcelona Supercomputer Center, through allocation of computer time of the Red Española de Supercomputación, and the Centro de Computación Científica de la Universidad Autónoma de Madrid. DD, FH and ER gratefully acknowledge fruitful discussions with David Gauthier (CEA-Saclay). M. M. acknowledges support by the Deutsche Forschungsgemeinschaft (DFG) under Grants No. SFB925/A3 and CUI, No. DFG-EXC1074. ER has been partially supported by the LABEX CEMPI (ANR-11-LABX-0007) as well as by the Ministry of Higher Education and Research, Hauts de France council and European Regional Development Fund (ERDF) through the Contrat de Projets Etat-Region (CPER) Photonics for Society (P4S). The authors gratefully acknowledge the machine physicists of FERMI who made this experiment possible by providing high-quality FEL light.

## References

1. Zewail, A. H. Femtochemistry. *J. Phys. Chem.* **97**, 12427–12446 (1993).
2. Nisoli, M., Decleva, P., Calegari, F., Palacios, A. & Martín, F. Attosecond Electron Dynamics in Molecules. *Chem. Rev.* **117**, 10760–10825 (2017).
3. Borrego-Varillas, R., Lucchini, M. & Nisoli, M. Attosecond spectroscopy for the investigation of ultrafast dynamics in atomic, molecular and solid-state physics. *Rep. Prog. Phys.* **85**, 066401 (2022).
4. Calegari, F. & Martin, F. Open questions in attochemistry. *Commun. Chem.* **6**, 184 (2023).
5. Kling, M. F. *et al.* Control of Electron Localization in Molecular Dissociation. *Science* **312**, 246–248 (2006).
6. Kling, M. F. & Vrakking, M. J. J. Attosecond Electron Dynamics. *Annu. Rev. Phys. Chem.* **59**, 463–492 (2008).
7. Prince, K. C. *et al.* Coherent control with a short-wavelength free-electron laser. *Nat. Photonics* **10**, 176–179 (2016).
8. Gauthier, D. *et al.* Generation of Phase-Locked Pulses from a Seeded Free-Electron Laser. *Phys. Rev. Lett.* **116**, 024801 (2016).
9. Usenko, S. *et al.* Attosecond interferometry with self-amplified spontaneous emission of a free-electron laser. *Nat. Commun.* **8**, 15626 (2017).
10. Giannessi, L. *et al.* Coherent control schemes for the photoionization of neon and helium in the Extreme Ultraviolet spectral region. *Sci. Rep.* **8**, 7774 (2018).
11. Gryzlova, E. V., Grum-Grzhimailo, A. N., Staroselskaya, E. I., Douguet, N. & Bartschat, K. Quantum coherent control of the photoelectron angular distribution in bichromatic-field ionization of atomic neon. *Phys. Rev. A* **97**, 013420 (2018).
12. Di Fraia, M. *et al.* Complete Characterization of Phase and Amplitude of Bichromatic Extreme Ultraviolet Light. *Phys. Rev. Lett.* **123**, 213904 (2019).

13. Allaria, E. *et al.* Highly coherent and stable pulses from the FERMI seeded free-electron laser in the extreme ultraviolet. *Nat. Photonics* **6**, 699–704 (2012).

14. Yin, Y.-Y., Chen, C., Elliott, D. S. & Smith, A. V. Asymmetric photoelectron angular distributions from interfering photoionization processes. *Phys. Rev. Lett.* **69**, 2353–2356 (1992).

15. Wang, Z.-M. & Elliott, D. S. Determination of the Phase Difference between Even and Odd Continuum Wave Functions in Atoms through Quantum Interference Measurements. *Phys. Rev. Lett.* **87**, 173001 (2001).

16. Baranova, N. B., Chudinov, A. N. & Zel'dovich, B. Ya. Polar asymmetry of photoionization by a field with $\langle E3 \rangle \neq 0$. Theory and experiment. *Opt. Commun.* **79**, 116–120 (1990).

17. Žitnik, M. *et al.* Interference of two-photon transitions induced by XUV light. *Optica* **9**, 692–700 (2022).

18. You, D. *et al.* New Method for Measuring Angle-Resolved Phases in Photoemission. *Phys. Rev. X* **10**, 031070 (2020).

19. Gryzlova, E. V. *et al.* Influence of an atomic resonance on the coherent control of the photoionization process. *Phys. Rev. Res.* **4**, 033231 (2022).

20. Maroju, P. K. *et al.* Attosecond coherent control of electronic wave packets in two-colour photoionization using a novel timing tool for seeded free-electron laser. *Nat. Photonics* **17**, 200–207 (2023).

21. Haessler, S. *et al.* Phase-resolved attosecond near-threshold photoionization of molecular nitrogen. *Phys. Rev. A* **80**, 011404 (2009).

22. Cattaneo, L. *et al.* Attosecond coupled electron and nuclear dynamics in dissociative ionization of H2. *Nat. Phys.* **14**, 733–738 (2018).

23. Nandi, S. *et al.* Attosecond timing of electron emission from a molecular shape resonance. *Sci. Adv.* **6**, eaba7762.

24. Bello, R. Y., Martín, F. & Palacios, A. Attosecond laser control of photoelectron angular distributions in XUV-induced ionization of $H_2$. *Faraday Discuss.* **228**, 378–393 (2021).

25. Suñer-Rubio, A. J. *et al.* Attosecond photoionization delays in molecules: The role of nuclear motion. *Phys. Rev. Res.* **6**, L022066 (2024).

26. Pollard, J. E., Trevor, D. J., Reutt, J. E., Lee, Y. T. & Shirley, D. A. Rotationally resolved photoelectron spectroscopy of *n* -H2, *p* -H2, HD, and D2. *J. Chem. Phys.* **77**, 34–46 (1982).

27. Holzmeier, F. *et al.* Control of H2 Dissociative Ionization in the Nonlinear Regime Using Vacuum Ultraviolet Free-Electron Laser Pulses. *Phys. Rev. Lett.* **121**, 103002 (2018).

28. Svetina, C. *et al.* The Low Density Matter (LDM) beamline at FERMI: optical layout and first commissioning. *J. Synchrotron Radiat.* **22**, 538–543 (2015).

29. Lyamayev, V. *et al.* A modular end-station for atomic, molecular, and cluster science at the low density matter beamline of FERMI@Elettra. *J. Phys. B At. Mol. Opt. Phys.* **46**, 164007 (2013).

30. Palacios, A., Feist, J., González-Castrillo, A., Sanz-Vicario, J. L. & Martín, F. Autoionization of Molecular Hydrogen: Where do the Fano Lineshapes Go? *ChemPhysChem* **14**, 1456–1463 (2013).

31. Sánchez, I. & Martín, F. The doubly excited states of the H2 molecule. *J. Chem. Phys.* **106**, 7720–7730 (1997).

32. Ito, K., Hall, R. I. & Ukai, M. Dissociative photoionization of H2 and D2 in the energy region of 25–45 eV. *J. Chem. Phys.* **104**, 8449–8457 (1996).

33. Palacios, A., González-Castrillo, A. & Martín, F. Molecular interferometer to decode attosecond electron–nuclear dynamics. *Proc. Natl. Acad. Sci.* **111**, 3973–3978 (2014).

34. Åsbrink, L. The photoelectron spectrum of $H_2$. *Chem. Phys. Lett.* **7**, 549–552 (1970).

35. Philip, J. *et al.* Highly accurate transition frequencies in the $H_2$ Lyman and Werner absorption bands. *Can. J. Chem.* **82**, 713–722 (2004).

36. Palacios, A., Bachau, H. & Martín, F. Enhancement and Control of $H_2$ Dissociative Ionization by Femtosecond VUV Laser Pulses. *Phys. Rev. Lett.* **96**, 143001 (2006).

37. Carpeggiani, P. A. *et al.* Disclosing intrinsic molecular dynamics on the 1-fs scale through extreme-ultraviolet pump-probe measurements. *Phys. Rev. A* **89**, 023420 (2014).

38. Palacios, A., Bachau, H. & Martín, F. Excitation and ionization of molecular hydrogen by ultrashort vuv laser pulses. *Phys. Rev. A* **75**, 013408 (2007).

39. González-Castrillo, A., Martín, F. & Palacios, A. Quantum state holography to reconstruct the molecular wave packet using an attosecond XUV–XUV pump-probe technique. *Sci. Rep.* **10**, 12981 (2020).

40. Gardner, J. L. & Samson, J. A. R. Proton kinetic-energy distributions from dissociative photoionization of hydrogen. *Phys. Rev. A* **12**, 1404–1407 (1975).

41. Lafosse, A. *et al.* Molecular frame photoelectron angular distributions in dissociative photoionization of $H_2$ in the region of the $Q_1$ and $Q_2$ doubly excited states. *J. Phys. B At. Mol. Opt. Phys.* **36**, 4683–4702 (2003).

42. Grum-Grzhimailo, A. N., Gryzlova, E. V., Staroselskaya, E. I., Venzke, J. & Bartschat, K. Interfering one-photon and two-photon ionization by femtosecond VUV pulses in the region of an intermediate resonance. *Phys. Rev. A* **91**, 063418 (2015).

43. Garcia, G. A., Nahon, L. & Powis, I. Two-dimensional charged particle image inversion using a polar basis function expansion. *Rev. Sci. Instrum.* **75**, 4989–4996 (2004).

44. Dowek, D. & Decleva, P. Trends in angle-resolved molecular photoelectron spectroscopy. *Phys. Chem. Chem. Phys.* **24**, 24614–24654 (2022).

45. Di Fraia, M. *et al.* Impulsive laser-induced alignment of OCS molecules at FERMI. *Phys. Chem. Chem. Phys.* **19**, 19733–19739 (2017).

46. Varvarezos, L. *et al.* Controlling Fragmentation of the Acetylene Cation in the Vacuum Ultraviolet via Transient Molecular Alignment. *J. Phys. Chem. Lett.* **14**, 24–31 (2023).

47. Decking, W. *et al.* A MHz-repetition-rate hard X-ray free-electron laser driven by a superconducting linear accelerator. *Nat. Photonics* **14**, 391–397 (2020).

48. Inoue, I. *et al.* Nanofocused attosecond hard x-ray free-electron laser with intensity exceeding $10^{19}$ W/cm$^2$. *Optica* **12**, 309 (2025).

49. Liu, T. *et al.* Status and future of the soft X-ray free-electron laser beamline at the SHINE. *Front. Phys.* **11**, 1172368 (2023).

50. Schoenlein, R., Boutet, S., Minitti, M. & Dunne, A. M. The Linac Coherent Light Source: Recent Developments and Future Plans. *Appl. Sci.* **7**, 850 (2017).

51. Sobolev, E. *et al.* Data reduction activities at European XFEL: early results. *Front. Phys.* **12**, 1331329 (2024).

52. Guo, Z. *et al.* Experimental demonstration of attosecond pump–probe spectroscopy with an X-ray free-electron laser. *Nat. Photonics* **18**, 691–697 (2024).

53. L A A Nikolopoulos & P Lambropoulos. Multichannel theory of two-photon single and double ionization of helium. *J. Phys. B At. Mol. Opt. Phys.* **34**, 545 (2001).

54. Foumouo, E., Kamta, G. L., Edah, G. & Piraux, B. Theory of multiphoton single and double ionization of two-electron atomic systems driven by short-wavelength electric fields: An ab initio treatment. *Phys. Rev. A* **74**, 063409 (2006).

55. Sánchez, I. & Martín, F. Resonant dissociative photoionization of H 2 and D 2. *Phys. Rev. A* **57**, 1006–1017 (1998).

56. Palacios, A., Sanz-Vicario, J. L. & Martín, F. Theoretical methods for attosecond electron and nuclear dynamics: applications to the H2 molecule. *J. Phys. B At. Mol. Opt. Phys.* **48**, 242001 (2015).

# Supplementary Material:

# Two-colour coherent control of nuclear and electron dynamics in photoionization of molecular hydrogen with FEL pulses

**Authors:** F. Holzmeier[1,2,*], A. Gonzalez-Castrillo[3,4], T. M. Baumann[5], R. Y. Bello[6], C. Callegari[7], M. Di Fraia[7], M. Lucchini[2,8], M. Meyer[5], O. Plekan[7], K. C. Prince[7,9], E. Roussel[10], R. Wagner[5], F. Martín[3,4,*], A. Palacios[3,11,*], D. Dowek[1,*]

**Affiliations:**

[1]Université Paris-Saclay, CNRS, Institut des Sciences Moléculaires d'Orsay, 91405 Orsay, France.

[2]Dipartimento di Fisica, Politecnico di Milano, 20133 Milan, Italy

[3]Departamento de Química, Universidad Autónoma de Madrid (UAM), 28049 Madrid, Spain

[4]Instituto Madrileño de Estudios Avanzados en Nanociencia (IMDEA-Nano), Campus de Cantoblanco, 28049 Madrid, Spain

[5]European XFEL, 22869 Schenefeld, Germany

[6]Departamento de Química Física Aplicada, UAM, 28049 Madrid, Spain

[7]Elettra-Sincrotrone Trieste, 34149 Basovizza, Italy

[8]Institute for Photonics and Nanotechnologies, IFN-CNR, 20133 Milano, Italy

[9]Department of Surface and Plasma Science, Charles University, Prague 18000, Czech Republic

[10]Université Lille, CNRS, UMR 8523, PhLAM-Physique des Lasers Atomes et Molécules, 59000 Lille, France

[11]Condensed Matter Physics Center (IFIMAC), Universidad Autónoma de Madrid (UAM), 28049 Madrid, Spain

*Correspondence to: fabian.holzmeier@imec.be, fernando.martin@uam.es, alicia.palacios@uam.es, danielle.dowek@universite-paris-saclay.fr

**Figure S1:** Franck Condon overlap and factors

(a) Franck Condon overlap of the $H_2(B\ ^1\Sigma_u^+, v'=6)$ neutral state and the $H_2^+(X\ ^2\Sigma_g^+, v_f)$ cationic ground state nuclear wavefunctions: $F(v',v_f) = \int dR\, \chi_{v'}(R)\chi_{v_f}(R)$

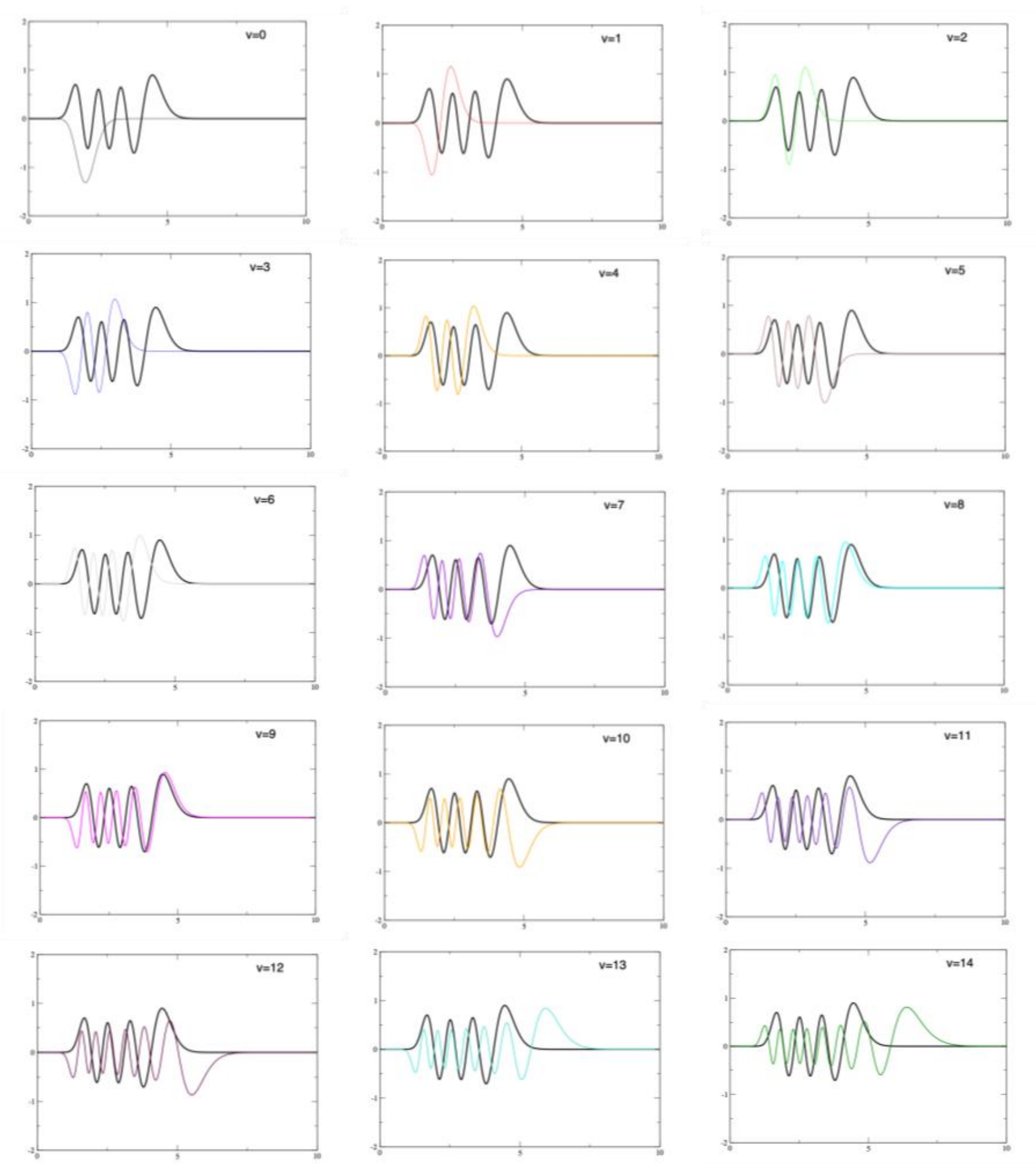

(b) Corresponding Franck Condon factors: $\left|F(v',v_f)\right|^2$

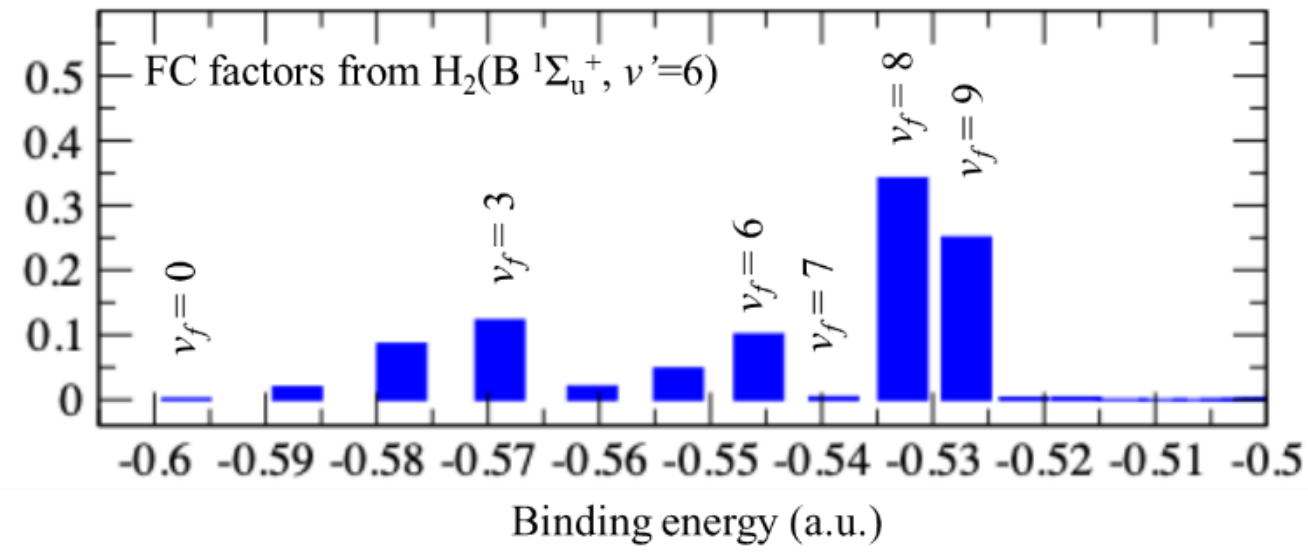

**Figure S2:** Computed photoelectron spectra

(a)

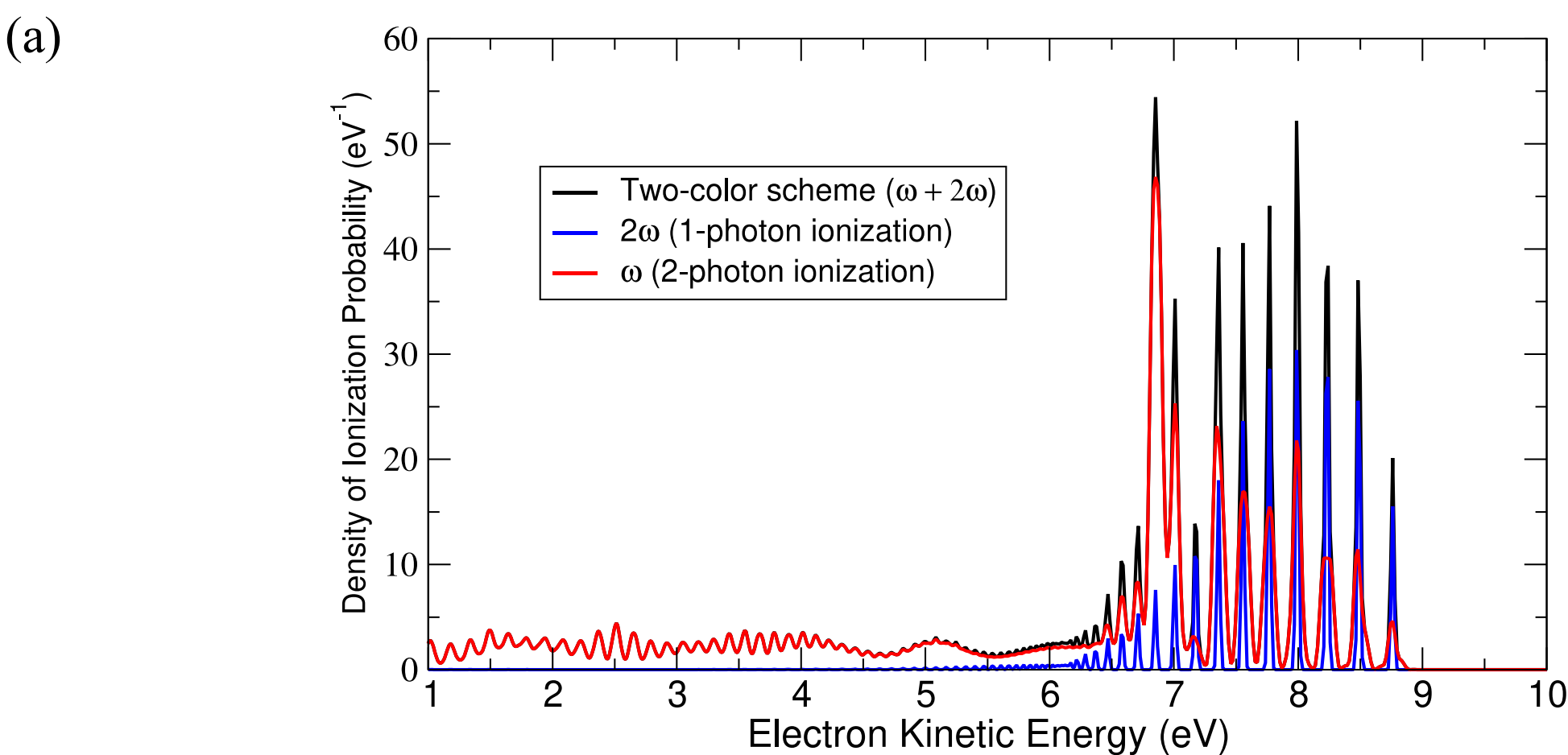

**Theoretical Photoelectron spectra for the one-photon and two-photon ionization pathways with no energy convolution.** We plot the density of ionization probabilities for the one-photon (blue curve), two-photon (red curve), and the ω-2ω scheme (black curve) for ionization of $H_2$ with $\hbar\omega$=12.1 eV and $\hbar 2\omega$ = 24.2 eV. It corresponds to the calculated spectra with an electronic energy grid using 1536 points in the interval of energies shown in the figure. We can see the well-defined vibrational lines of the bound states (higher electron energies). The fast oscillations in the lower part of the spectra are a numerical artefact due to the employment of very long pulses (100 fs). As is well known, a much finer discretization would be required for a fully smooth background. Nevertheless, convergence is reached (we realized the calculation with an increasing number of grid points from 300 up to 1586 points). Laser peak intensities in the calculation were chosen as $2.5 \times 10^{13}$ and $10^{12}$ W/cm$^2$, for the 2ω and ω FEL radiations, respectively.

(b)

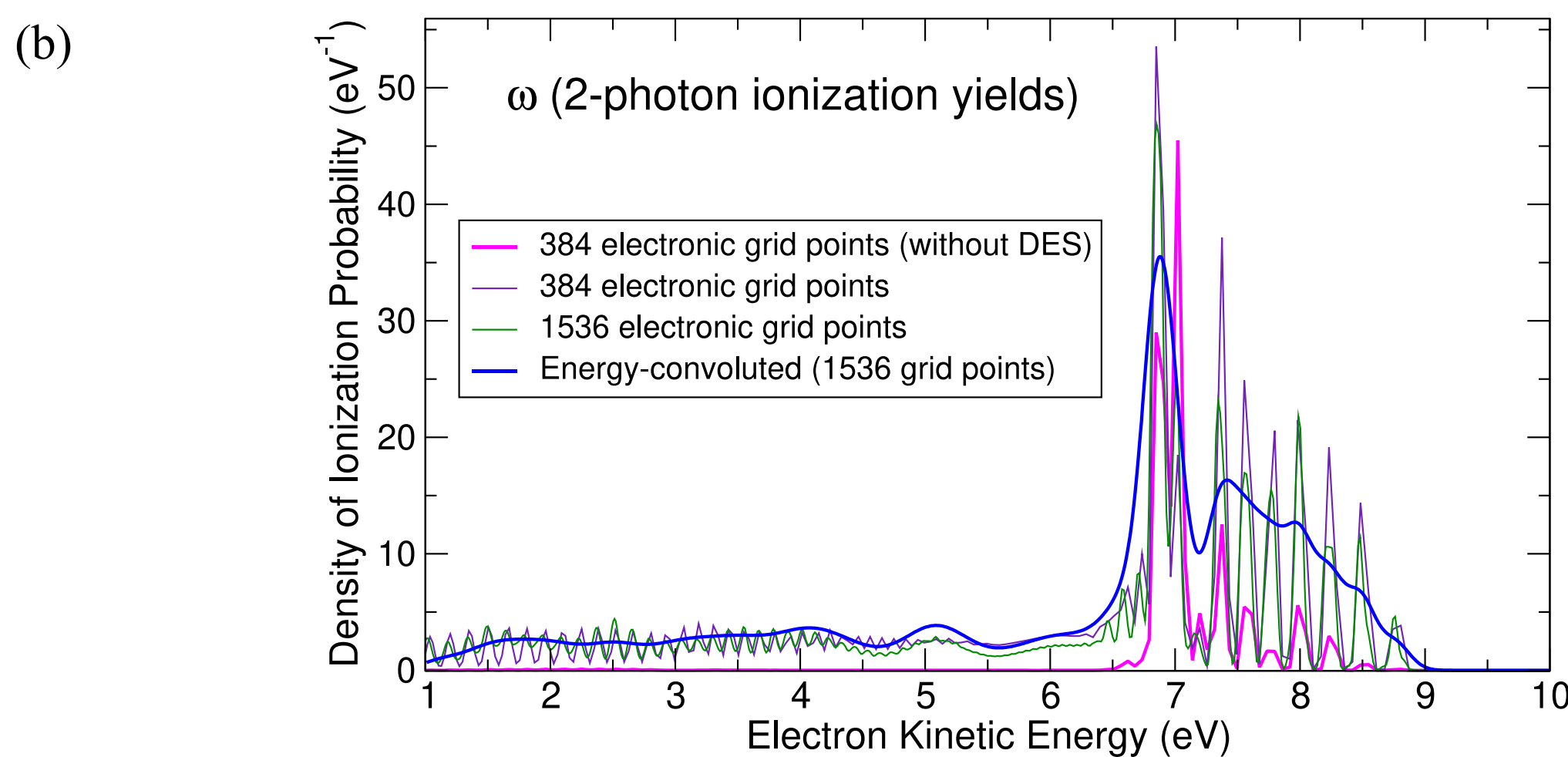

**Theoretical Photoelectron spectra (density of ionization probabilities) for the two-photon ionization pathway** with $\hbar\omega$=12.1 eV, a pulse duration of 100 fs and an intensity of $10^{12}$ W/cm$^2$. We first show the numerical convergence with the number of electronic grid points within the energy range shown in the figure (violet thin line with 384 and green thin line with 1536 grid points). We can see that the vibrational peaks become more "defined" in energy as the number of points increases (higher density of the discretized electronic continuum states). The blue thick line is the energy-convoluted curve shown in Fig. 3b (red full line) as explained in the main manuscript. The magenta thick line corresponds to the results removing the doubly excited states, i.e., removing the possibility of autoionization upon two-photon absorption. We can see that the vibrational progression that is captured in the photoelectron spectra for the two-photon path faithfully mirrors the Franck-Condon factors given in Figure 3d in the main manuscript.

**Figure S3:** Amplitude $B(\theta_k, \varepsilon)$of the oscillatory term of the PAD $I(\theta_k, \phi, \varepsilon)$ (Eq.1)

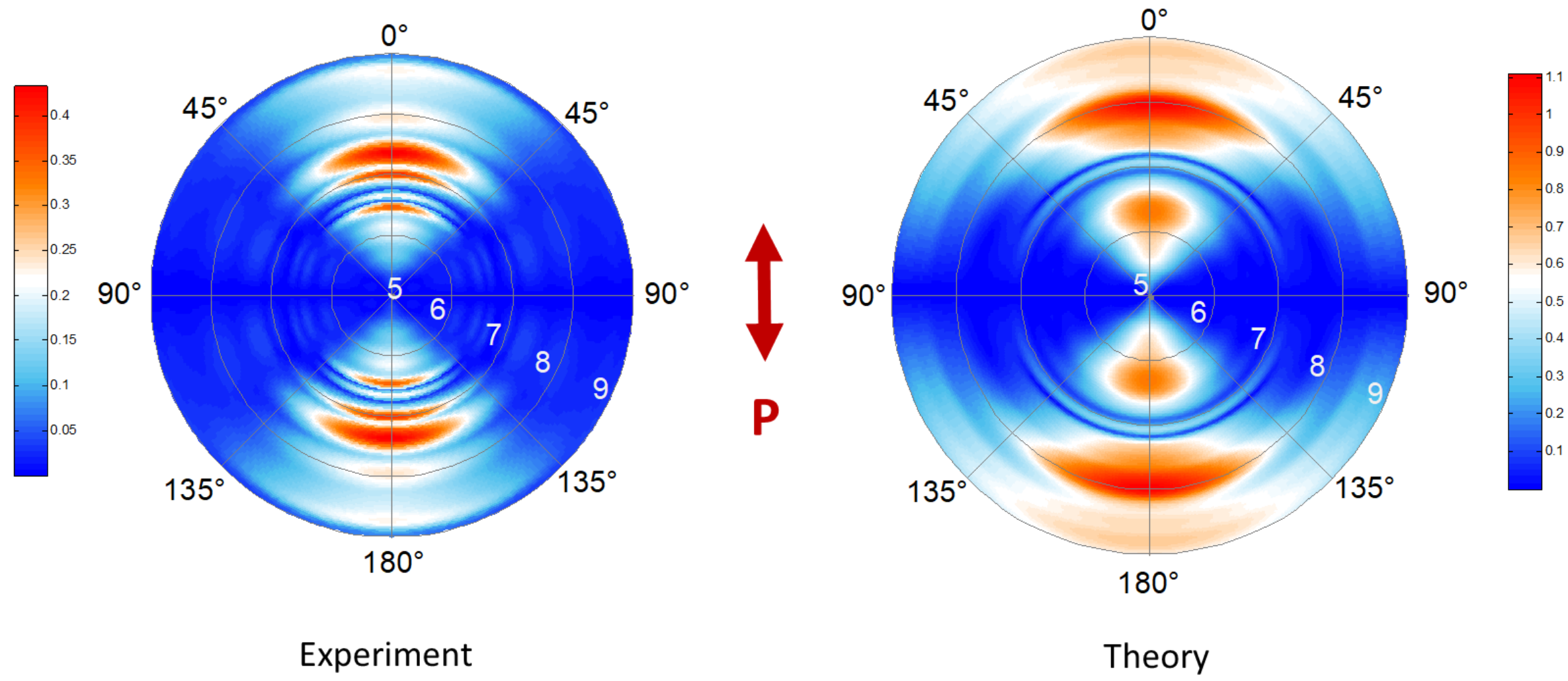

The polar emission angle $\theta_k$ is defined relative to the polarization axis of the FERMI radiation P. The photoelectron energy $\varepsilon$ varies between 5 and 9 eV along the radius of the 2D plot.